\newif\iffigs\figstrue

\iffigs
   \documentstyle[12pt,epsf]{article}
\else
   \documentstyle[12pt]{article}
\fi

\def\thefootnote{\fnsymbol{footnote}}

\newcommand{\eq}{\begin{equation}}
\newcommand{\en}{\end{equation}}
\newcommand{\eqa}{\begin{eqnarray}}
\newcommand{\ena}{\end{eqnarray}}

\newcommand{\lbl}{\label}

\newcommand{\bhk}{J^H}
\newcommand{\bw}{J_w}
\newcommand{\ca}{{\cal A}}

\newcommand{\NP}[1]{Nucl.\ Phys.\ {\bf #1}}
\newcommand{\PL}[1]{Phys.\ Lett.\ {\bf #1}}

\newcommand{\PR}[1]{Phys.\ Rev.\ {\bf #1}}
\newcommand{\PRL}[1]{Phys.\ Rev.\ Lett.\ {\bf #1}}

\begin{document}

\iffigs
  \input epsf
\else
  \message{No figures will be included. See TeX file for more
information.}
\fi

\hskip 11.5cm \vbox{\hbox{DFTT 30/94}\hbox{SISSA 103/94/EP}\hbox{July 1994}}
\vskip 0.4cm
\centerline{\Large\bf Deconfinement Transition in Large $N$}
\centerline{\Large\bf Lattice Gauge Theory}
\vskip 1cm
\centerline{M. Bill\'o \footnote{billo@tsmi19.sissa.it}}
\vskip .3cm
\centerline{\sl SISSA, Via Beirut 2-4, I-34013, Trieste, Italy}
\centerline{\sl Istituto Nazionale di Fisica Nucleare, 
Sezione di Trieste}
\vskip .5cm
\centerline{M. Caselle, A. D'Adda, L. Magnea
\footnote{caselle@to.infn.it}}
\vskip .3cm
\centerline{\sl Istituto Nazionale di Fisica Nucleare, Sezione di Torino}
\centerline{\sl  Dipartimento di Fisica 
Teorica dell'Universit\`a di Torino}
\centerline{\sl via P.Giuria 1, I-10125 Turin,Italy}
\vskip .5cm
\centerline{S. Panzeri}
\vskip .3cm
\centerline{\sl SISSA,Via Beirut 2-4, I-34013, Trieste, Italy}
\centerline{\sl Istituto Nazionale di Fisica Nucleare, 
Sezione di Torino}

\vskip 1.5cm

\begin{abstract}

We study analytically the phase diagram of the pure $SU(N)$ lattice
gauge theory at finite temperature, and we attempt to estimate the 
critical deconfinement temperature. We apply large $N$ techniques to the 
Wilson  and to the Heat Kernel action, 
and we study the resulting models both in the strong coupling and in the 
weak coupling limits. Using the Heat Kernel action, we establish an 
interesting connection between the Douglas-Kazakov phase transition of 
two-dimensional QCD and the deconfining transition in $d$ dimensions.
The analytic results obtained for the critical temperature compare well 
with Montecarlo simulations of the full theory in $(2+1)$ and in $(3+1)$ 
dimensions.

\end{abstract}
\noindent
\vfill
\eject

\newpage

\setcounter{footnote}{0}
\def\thefootnote{\arabic{footnote}}

\section{Introduction}

Finding a precise characterization of the finite temperature deconfining 
transition is a long standing problem of Lattice Gauge Theory (LGT). 
Since the seminal work of Svetitsky and Yaffe~\cite{sy}, it 
is understood that all the relevant properties of the transition can be  
encoded in a suitable effective action for the order parameter, the 
Polyakov loop, and a qualitatively
satisfactory description of the phase diagram can 
be obtained in this way. Such an approach clearly
involves an enormous simplification of the model, through a drastic 
reduction of its degrees of freedom;  the non-trivial task is then 
the construction of an appropriate effective action. 
The simplest and most popular proposal for such an action 
is~\cite{og}-\cite{dh}
\eq
S_{eff}(J)=\sum_{\vec x}~Re\left\{J\sum_{i}
~{\rm Tr_f}(V(\vec x)){\rm Tr_f}(V^\dagger(\vec{x}+i))\right\}~~~,
\label{seff}
\en
where ${\rm Tr_f}(V(\vec x))$ is the Polyakov loop.

This action has been widely studied in the literature, 
both at finite $N$~\cite{og,djk} and in the large $N$ 
limit~\cite{dp,gnr}, with strong coupling~\cite{ps,gk} and 
mean field~\cite{djk} approximations,
and with Montecarlo simulations~\cite{djk,dh}. 
In particular, in the large $N$ limit it is exactly solvable, 
and leads to a peculiar phase diagram, with a first order point and 
a line of third order transitions of the Gross-Witten 
type~\cite{dp,gnr}.
In the literature, the action (\ref{seff}) has been mainly used as 
a tool to understand the general features of the phase diagram,
without really hoping to obtain precise 
determinations of, say, the location of the critical deconfinement 
temperature $T_c$. This is due to the fact that the approximations 
involved in going from the complete LGT to eq. (\ref{seff}) are rather 
drastic, and the correspondence between the coupling $J$ in 
eq. (\ref{seff}) and the coupling $\beta$ of 
the complete LGT (and the consequent possibility of 
finding a meaningful result for the critical temperature in the 
continuum limit) is rather ambiguous.

One of the goals of the present paper is to show that, by using 
new techniques which have become available in these last years 
for the large $N$ analysis of gauge theories,
one can go a bit further and extract from a 
suitable generalization of the action (\ref{seff}) some reliable 
information on the deconfinement temperature. 
Even if the values obtained are still 
affected by large uncertainties, they essentially agree with those
extracted from Montecarlo simulations. This is a significant progress,
because the results are obtained analitycally and, what is more 
important, our approach opens the way to further improvement.
In fact a better control on the approximations involved may be obtained 
within the framework of the hot Eguchi-Kawai model, and also the 
extension to the case of finite $N$ may be studied.

In this paper, we shall show that the action (\ref{seff}) can 
be considered as an approximation of a much more accurate effective 
action which, despite its complexity, can still be precisely studied in 
the large $N$ limit, in weak coupling as well as in strong coupling. 
Besides describing the phase diagram of the model, we will make an 
effort to extract as precisely as possible the value of the critical 
temperature in terms of the coupling of the original model, studying 
its scaling behaviour and making contact with the known results from 
Montecarlo simulations both in (2+1) and (3+1) dimensions. 
As we shall see in the last section, 
the two cases require very different analysis, and 
only in the (2+1)-dimensional case the approximations are 
really under control, the value that we obtain has the correct 
scaling behaviour, and the continuum limit can be correctly approached.
In (3+1) dimensions things are significantly more complicated, 
and our results should be better taken as strong coupling estimates, valid 
for small values of the $N_t$ (the lattice size in the time direction); 
nevertheless, also in this case, some interesting information can be 
extracted, and a rather succesfull comparison with known Montecarlo 
estimates can be made. 

To describe QCD on the lattice, we will use both the Wilson action and
the action defined in terms of the Heat Kernel on the group 
manifold~\cite{heatkernel}, already employed in this context 
by~\cite{zar}. We will study the approximations that are needed 
to reduce these actions to the model (\ref{seff}), namely a 
Migdal-Kadanoff renormalization and the decoupling of the space-like 
plaquettes. We observe that the Heat Kernel action 
has definite advantages from the point of view of analytic calculations,
 and, upon 
decoupling the space-like plaquettes, allows an exact integration over 
the space-like degrees of freedom. 
The resulting model 
can be studied, in the large $N$ limit, both in weak coupling (where it 
is directly equivalent to the Wilson action), and in strong coupling, 
where a non-trivial transformation is needed to make the connection
with the Wilson action.

In the process, we shall make contact with some 
interesting result recently obtained in the context of two-dimensional 
QCD (QCD2). In particular, we shall discuss the relation between the 
deconfinement transition in $(d+1)$ LGT and the so called 
Douglas-Kazakov transition~\cite{dk} in  QCD2. 

Our analysis starts with a review of the various approximations needed to 
obtain $S_{eff}$. This will be done in the next section, where we shall 
also fix our notations in the general framework of finite temperature 
LGT. Section 3 will be devoted to a discussion of how the same 
approximations, applied to the Heat Kernel, link the effective action 
for the Polyakov loop to the  two-dimensional QCD on a cylinder.
We then 
review some of the features of QCD2 which are relevant in this case. 
Sections 4 and 5 will be devoted to a weak coupling and a strong 
coupling expansion respectively, while in the last section we shall  
make contact with the known results on the deconfinement temperature 
of LGT, and we shall make some concluding remarks. 

\section{Finite Temperature LGT}

Let us consider a pure gauge theory with gauge group $SU(N)$, defined 
on a $d+1$ dimensional cubic lattice. 
In order to describe a finite temperature LGT,
we have to impose periodic 
boundary conditions in one direction (which we shall call from now 
on ``time-like'' direction), while the boundary conditions in the 
other $d$ direction (which we shall call ``space-like'') can be 
chosen freely. 
We take a lattice of $N_t$ ($N_s$) spacings in the time (space) 
direction, and we work with the pure gauge theory, containing
only gauge fields described by the link 
variables $U_{n;i} \in SU(N)$, where $ n \equiv (\vec x,t)$ 
denotes the space-time position of the link and $i$ its direction. 
It is useful to choose different bare couplings in the time and space 
directions. Let us call them $\beta_t$ and $\beta_s$ respectively. 
The Wilson action is then
\eq
S_W=\sum_{n}~\frac{1}{N}~Re\left\{\beta_t\sum_i~{\rm Tr_f}(U_{n;0i})
+\beta_s\sum_{i<j}~{\rm Tr_f}(U_{n;ij})\right\}~~~,
\label{wilson}
\en
where ${\rm Tr_f}$ denotes the trace in the fundamental representation
and $U_{n;0i}$ ($U_{n;ij}$) are the time-like (space-like) 
plaquette variables, defined as usual by
\eq
U_{n;ij}=U_{n;i}U_{n + \hat i;j}
U^\dagger_{n + \hat j;i}U^\dagger_{n;j}~~~.
\en

In the following we shall call $S_s$ ($S_t$) the space-like (time-like) 
part of $S_W$. $\beta_s$ and $\beta_t$ are related to the (bare) gauge 
coupling $g$ and to the temperature $T$ by the usual relations

\eq
\frac{2N}{g^2}=a^{3-d}\sqrt{\beta_s\beta_t}~~~,
\hskip 1cm 
T=\frac{1}{N_ta}\sqrt{\frac{\beta_t}{\beta_s}}~~~,
\label{couplings}
\en
where $a$ is the space-like lattice spacing, while $\frac{1}{N_tT}$ is
the time-like spacing. The two are related by the dimensionless ratio 
$\epsilon\equiv \frac{1}{N_tTa}$. 
We can solve the above equations in terms of $\epsilon$, as
\eq
\beta_t=\frac{2N}{g^2\epsilon}a^{d-3}~~~,
\label{betat}
\en
\eq
\beta_s=\frac{2N\epsilon}{g^2}a^{d-3}~~~.
\label{betas}
\en

In a finite temperature discretization it is possible to define 
gauge invariant observables which are topologically non-trivial, 
as a consequence of the periodic boundary conditions in the 
time directions. 
The simplest choice is the Polyakov loop, 
defined in terms of link 
variables as
\eq
P(\vec x)\equiv 
{\rm Tr_f}~V(\vec x)= {\rm Tr_f} \prod_{t=1}^{N_t}(U_{\vec x,t;0})~~~.
\label{polya}
\en

As it is well known, the finite temperature theory has a new 
global symmetry (unrelated to the gauge symmetry), with symmetry group 
the center $C$ of the gauge group (in our case $Z_N$). The Polyakov loop 
is a natural order parameter for this symmetry.

In $d>1$, finite temperature gauge theories admit a 
deconfinement transition at $T=T_c$, separating the
high temperature, deconfined, phase ($T>T_c$) from the low 
temperature, confining domain ($T<T_c$). 
In the following we shall be interested in the phase diagram of the 
model as a function of $T$, and shall make some attempt to locate the 
critical point $T_c$. The high 
temperature regime is characterized by the  breaking of the global 
symmetry with respect of the center of the group. In this phase the 
Polyakov loop has a non-zero expectation value, and it is 
an element of the center of the gauge group 
(see for instance~\cite{sy}). 

The Wilson action eq.(\ref{wilson}) is clearly too complex to be 
handled exactly, and some approximate effective action is needed.
As we shall see, at least three different approximations are needed, in 
order to reach a solvable model. 
The first step is to identify the Polyakov loops as relevant dynamical 
variables. This can be done by dimensional reduction, 
as described in the next subsection (2.1). Second, in order to
integrate out the spacelike degrees of freedom, 
one must decouple the spacelike plaquettes, and this will be
described in subsection (2.2). The third and last step is to take the 
large $N$ limit, which allows to solve exactly the model, through a large 
$N$ factorization of the Polyakov loops (subsection 2.3); the large $N$ 
limit is important even when it does not lead to exact solvability, 
because in that limit a translational invariant master field
is expected, and 
the model can be reduced to an effective one-plaquette model.

In the remaining part of this section we shall describe in some more 
detail these steps, and discuss their reliability. 
Since the approximations are well known, and have been already 
widely discussed in the literature, we shall mostly 
state the results, and refer the reader to some of
the original papers. 

\subsection{Migdal-Kadanoff approximation}
By means of an approximate renormalization group transformation, within 
the framework of a Migdal-Kadanoff bond-moving scheme, it is possible to 
reduce the original $(d+1)$ dimensional Wilson action, 
eq. (\ref{wilson}) to a $d$ dimensional gauge theory coupled to the 
Polyakov lines, which play the role of a Higgs field~\cite{og,djk}. 
The resulting action is
\eqa
\exp(S_1) & = & \prod_{\vec x}\left[\prod_{i=1}^d\left\{1+\sum_{r\neq 0}
\left[\frac{I_r(\beta_t)}{I_0(\beta_t)}\right]^{N_t}
~d_r\chi_r(V(\vec x)U_{\vec x;i}V^\dagger(\vec{x}+i)
U^\dagger_{\vec x;i})\right\}\right]
\times \nonumber \\
& \times & \exp\left[\frac{1}{N}~Re
\left\{\hat\beta_s\sum_{i<j}~{\rm Tr_f}(U_{\vec x;ij})\right\}
\right]~~~.
\label{ht}
\ena
Here the sum is limited to the $d$ spacelike 
directions, $I_r(\beta_t)$ is the coefficient of 
representation $r$ in the character expansion of the Wilson action,
and the new spacelike coupling is related to the old one by

\eq
\hat\beta_s=N_t\beta_s~~~.
\en

In the literature the timelike part of this action is then further
approximated by truncating the sum over representations
at the first term, so as to reconstruct once more a 
Wilson-type action

\eq
S_2=\sum_{\vec x}~\frac{1}{N}~Re\left\{\hat\beta_t\sum_{i=1}^d
~{\rm Tr_f}(V(\vec x)U_{\vec x;i}V^\dagger(\vec{x}+i)
U^\dagger_{\vec x;i})
+\hat\beta_s\sum_{i<j}~{\rm Tr_f}(U_{\vec x;ij})\right\}~,
\label{ht2}
\en

As we shall see, this  truncation is not necessary if one works with
the Heat Kernel action, to be discussed in the next section. In the 
present framework, the Migdal-Kadanoff approximation
is expected to be a good approximation both at strong and at weak 
coupling (see for instance~\cite{kad})) while it might be inadequate 
at intermediate couplings. In particular, it is easy to see that 
the timelike part of $S_1$, 
eq. (\ref{ht}), could have been obtained also within the 
framework of a strong coupling expansion.

The truncation of eq. (\ref{ht}) to eq. (\ref{ht2}) is similarly
expected to be valid only for large or small values of $\beta_t$, 
the difference between the two regimes being only in the definition of 
the new timelike coupling 
\eq
\hat\beta_t \sim \frac{\beta_t}{N_t}~~,~~~~~~~\beta_t~~{\rm large}~~~,
\label{blarge}
\en
\eq
\hat\beta_t\sim\frac{\beta_t^{N_t}}{(2N^2)^{N_t-1}}~~
,~~~~~~~\beta_t~~{\rm small}~~~.
\label{bsmall}
\en

In the large $\beta_t$ limit the Migdal-Kadanoff approximation coincides
 with 
the approximation scheme for high temperature QCD known in the 
literature as
``complete dimensional reduction''.
This consists in the assumption of complete decoupling of the non-static 
modes in the compactified time direction.
The resulting 
 theory is a $d$-dimensional gauge theory coupled to a scalar field in 
the adjoint representation
which represents the fluctuations of the Polyakov loop around the minimum 
$V(\vec{x}) = 1$.
The corresponding 
action is the sum of the purely space-like part of $S_W$ and a new term
$S_h$ which is the remnant of the timelike part of 
$S_W$
\eq 
S_h(m_0)=\frac{\beta_h}{N}~{\rm Tr_f}\sum_{\vec x} 
\left(-m_0^2 \phi(\vec x)^2+
\sum_{i=1}^d U_{\vec x;i}\phi(\vec x)U^\dagger_{\vec x;i}
\phi(\vec x+\hat i)\right)
\label{higgs}
\en
where $\phi(x)$ is an Hermitian $N \times N$ matrix and
$m_0^2=d$.

This action is of the type studied by Kazakov and Migdal
in the context of ``induced QCD'' \cite{KazMig} and in the present context
by the authors of \cite{CDPht}. 
It is however well known
that complete dimensional reduction does not take place 
in general and that non-static modes  
induce at one loop  level
some well defined interaction  in the static sector.
We expect this type of corrections to be the main source of error also
in the Migdal-Kadanoff approximation, however it can be shown that they are 
of higher order in large $N$ limit and therefore we shall not take them
into account in the present paper. 

Let us conclude by noticing that the  effect of the Migdal-Kadanoff 
approximation is to reduce the lattice size in the time direction $N_t$ 
to its extreme value $N_t=1$. It will be important to keep in mind this 
fact in the last 
section, where we shall compare our prediction with the results of 
the Montecarlo simulations. 

\subsection{Decoupling of spacelike plaquettes}

An interesting feature of eq.(\ref{ht})
is that its timelike and spacelike sectors 
are only very weakly coupled. Let us try to make this statement 
 more precise. First of all one can see 
that in the framework 
of a standard mean field approximation the decoupling
of timelike and spacelike degrees of freedom in (\ref{ht}) 
is complete~\cite{djk}.
This result appears to be confirmed beyond the mean field 
approximation by montecarlo simulations~\cite{djk}.
The decoupling was also checked in~\cite{cd}, where the contribution 
of timelike degrees of freedom to the space-like string tension
at high temperature was evaluated in (2+1) dimensions within a strong 
coupling expansion and  found to be neglected.
This result turned out to be in agreement with 
high precision Montecarlo simulations~\cite{t2}. 

In this paper we are addressing the problem of 
finding the critical deconfinement temperature
and the relevant degrees of freedom appear 
only in  the 
timelike part of the action. So, in agreement with the 
arguments given above, and following the 
literature on the subject~\cite{gnr,dp,dh},
we shall from now on neglect the spacelike 
part of the action, eq.(\ref{ht}). 
It is of course far from obvious to what extent such a crude approximation
still holds at temperatures of the order of the 
deconfinement transition, so an independent check of its validity 
(such as results from Montecarlo simulations) is going to be important.

The main outcome of this approximations is that in the resulting model 
the spacelike degrees of freedom can be explicitly integrated out,
to obtain an effective action for the Polyakov loops only. 
This integration can be made (see section 5) by using a 
generalization of the Itzykson-Zuber-Harish-Chandra integral~\cite{izhc}.
If one keeps only the lowest order in the coupling one recovers the 
action $S_{eff}$, eq.(\ref{seff}) discussed in the introduction.

\eq
S_{eff}(J)=\sum_{\vec x}~Re\left\{J\sum_{i}
~{\rm Tr_f}(V(\vec x)){\rm Tr_f}(V^\dagger(\vec{x}+i))\right\}~~~.
\label{riseff}
\en
Here we have introduced the new coupling $J$  defined in terms of 
$\beta$ as:
\eq
J=\hat\beta_t/N^2~~~.
\label{jbeta}
\en
$J$ has a smooth large $N$ behaviour and so it is the natural 
coupling in the large $N$ limit. In the following two sections we shall 
sistematically use $J$, referring to eq.(\ref{jbeta}) for the connection 
with $\beta$, and shall come back to the original coupling $\beta$ only in 
the last section, where the comparison with the Montecarlo results is 
made.

Higher order corrections to eq. (\ref{riseff}) are discussed in Section 5
where a similar analysis for the Heat Kernel action is also carried out.
 As we shall see, a phase diagram  much richer than that 
of $S_{eff}$ emerges if one keeps the whole result of the integration. 

\subsection{Large N Limit}

The model described by $S_{eff}$ can be solved exactly in the large 
$N$ limit~\cite{dp}, for any value of the space dimensions $d$, 
leading to a phase diagram with a first order phase transition located 
at $J=1/d$.
It is even possible to solve exactly the more general model in which the 
Polyakov loop is also coupled to an external ``magnetic'' field $h$,

\eqa
S_{eff}(J,h) & = & \sum_{\vec x}~Re\left\{J\sum_{i}
~{\rm Tr_f}(V(\vec x)){\rm Tr_f}(V^\dagger(\vec{x}+i))\right\}
~~+ \nonumber \\
& + & hN\sum_{\vec x}\left[
{\rm Tr_f}(V(\vec x) + V^\dagger(\vec{x}))\right]~~~.
\label{sjh}
\ena

In this more general framework it becomes apparent that the first order 
phase transition is the end point of a line of third order phase 
transitions of the Gross-Witten type, located along the line
\eq
J=\frac{1-2h}{d}~,~~~h,J\geq 0~~~~~.
\en

\subsection{Comments}

There are two interesting observations which must be made at this 
point. First it is possible to show that the three 
steps of approximation described above {\sl commute}, 
in the sense that the order in which they are made 
can be changed, but the same phase diagram is found at the end. 
For instance, if we neglect from the beginning the spacelike plaquettes, 
then the Migdal Kadanoff result eq. (\ref{ht}) can be obtained exactly 
within a strong coupling expansion (namely integrating all the spacelike 
links except the ones in the lowest slice). Similarly one could start 
from the beginning within the framework of a large $N$ Eguchi-Kawai 
model, then make a Migdal Kadanoff approximation, and find again the 
same phase diagram (see~\cite{gnr} and this same observation 
in~\cite{dp}).
This denotes an internal consistency of the set of approximations used
which suggests that the value $J_c=1/d$ for the deconfinement temperature
should be taken seriously. Indeed, such value is rather stable under 
the various   refinements of eq. (\ref{seff}) that will be discussed
in the rest of this paper; at the end our results will never differ
from it more than 20\%.

A second observation is that the magnetic field introduced in 
eq.(\ref{sjh}) has a relevant physical meaning. It encodes the effect 
of the coupling of fermions to the pure gauge theory~\cite{bu,bhdg}. 
To be precise, one has to choose Wilson fermions and then make an 
expansion in the hopping 
parameter $K$. It is possible to see that at the lowest order 
in the $K$ expansion the $Z_N$ symmetry of the  pure gauge theory is 
broken and that the fermions contribute with a magnetic term 
$h=2N_f(2K)^{N_t}$, where $N_f$ is the number of fermions introduced.

\section{The Heat Kernel Action}

The Heat Kernel action is best defined in terms of the character 
expansion 
\eqa
\exp(S_{H})= & &\prod_{n,(i<j)}\left\{\sum_r 
d_r\chi_r(U_{n;ij})\exp\left(-\frac{N C^{(2)}_r}{2\beta_s^{H}}\right)
\right\}\times\nonumber \\ & &\times\prod_{n,i}\left\{\sum_r d_r\chi_r(U_{n;0i})
\exp\left(-\frac{N C^{(2)}_r}{2\beta_t^{H}}\right)\right\}
\label{hk}
\ena
where the sum is over all the inequivalent,
 irreducible representations, labelled by 
$r$, $d_r$ is their dimension, $\chi_r(U)$ is the character of $U$ in 
the representation $r$ and $C_r^{(2)}$ is the value of the 
quadratic Casimir operator in the representation $r$. The index $s$ and 
$t$ label, as before, the spatial an temporal couplings respectively.
The Heat Kernel action in eq.(\ref{hk}) is normalized so as to coincide 
in the continuum limit $(\beta_s,\beta_t \to \infty)$ with the Wilson 
action eq.(\ref{wilson}). Hence in this limit 
eqs.(\ref{couplings},\ref{betat},\ref{betas}) 
still hold with the substitution $\beta\to \beta^{H}$

In the case of the Wilson action, both the Migdal Kadanoff
approximation (section 2.1) and  the 
decoupling and subsequent integration of the spatial degrees of 
freedom (sect. 2.2) require a truncation of the resulting actions so
as to obtain the final form of the effective action eq.(\ref{sjh}).
The effect of these truncations can be neglected if $\beta_t$ is very
large or very small, 
but it is important in the intermediate region in which we expect the 
transition to occur. The situation is different for the Heat Kernel 
action. In fact due to the well known 
properties of the Heat Kernel action the strong coupling expansion which leads 
to eq.(\ref{ht}) is exact in the timelike direction and the scaling of 
the Heat Kernel coupling is trivial:
\eq
\beta^{H}_t\to \hat\beta^{H}_t=\frac{\beta^{H}_t}{N_t}
\label{bheatkernel}
\en
for all values of $\beta^{H}_t$ and no truncation needed. 
As mentioned above, near the continuum 
limit  (large $\beta$) the Heat Kernel and the Wilson action must 
coincide, in fact the same scaling law appears in eq. (\ref{blarge}) and
(\ref{bheatkernel}).
On the contrary in the strong coupling regime the two 
behave very differently (as can be seen by comparing 
eq.(\ref{bheatkernel}) and eq.(\ref{bsmall}) ). This can be  checked
by a direct comparison of the strong coupling expansion of 
the Wilson and the Heat Kernel action (see section 5).
The strong coupling expansion provides a relation between 
$\beta^H$ and $\beta$,which, as expected  compensates
exactly  the difference between the two scaling laws 
eqs. (\ref{bheatkernel}), (\ref{bsmall}). This not trivial relation must 
be kept into account if one wants to make contact with say, results 
coming from Montecarlo simulations which are always made with the 
standard Wilson action.

The decoupling of the spatial plaquettes, which we shall assume also in 
this context, makes it possible to integrate exactly on the spatial 
links by using standard properties of the characters. We have 
\eqa
&&Z_H =\nonumber\\
&& = 
 \int \prod_{\vec x} dV(\vec x) \prod_{\vec x,i} dU_i(\vec x) 
\prod_{\vec x,i} \left\{ \sum_r d_r \chi_r \left( U_i(\vec x)
 V(\vec x + i) U^{\dagger}_i(\vec x) V^{\dagger}(\vec x) \right) \exp 
\left( -\frac{C^{(2)}_r}{2 N J^H}\right)\right\}  \nonumber \\
& & = \int \prod_{\vec x} dV(\vec x)  \prod_{\vec x,i} \sum_r \chi_r \left
( V(\vec x + i) \right) \chi_r \left( V^{\dagger}(\vec x) \right) \exp
\left( -\frac{C^{(2)}_r}{2 N J^H}\right) \label{boh}
\ena
 
The effective action for the Polyakov loop given by (\ref{boh}), 
although  much more complicated than that of eq.(\ref{sjh}), can still be
studied in the large N limit. This was done, both in the weak and in the 
strong coupling region, by K. Zarembo in a recent paper 
~\cite{zar}.  A rather puzzling result of \cite{zar} is the 
logarithmic dependence on the dimensions of the critical point, more 
precisely 
\eq
J_c ={1 \over 2 \log(2 d - 1)}
\label{jzar} \en
This is a strong coupling result, obtained by looking at the point where 
the symmetric vacuum becomes unstable, and it is apparently in 
disagreement with the results of the previous section which give a $1/d$ 
behaviour for the critical coupling
Such disagreement is completely eliminated if one takes into account the 
relation between $\beta^{H}$ and $\beta$.

In the weak coupling region the action (\ref{boh}) can be studied by 
reducing it to a  Kazakov-Migdal model with a quadratic potential, whose 
solution is given by a semicircular distribution of eigenvalues for the 
Polyakov loop. In ref. ~\cite{zar} it was remarked that an instability 
in the solution appears for the value of the coupling constant where the 
radius of the distribution develops an imaginary part, presumably a 
signal of some type of phase transition.
Here we will produce a better understanding of this phenomenon, which 
will also allow a more precise determination of the critical coupling.
In order to do so we notice that the coupling between $V(\vec x)$ and
$V(\vec x + i)$ at 
the r.h.s of eq. (\ref{boh}) is just the partition function of QCD2 on 
cylinder whose area  
\footnote{As usual in the contest of QCD2,
the area ${\cal A}$ of the cylinder is expressed in units of 
$\frac{1}{g_2^2}$, where $g_2$ is the QCD2 coupling. }
${\cal A }$
is ${1 \over J^H}$ and whose fixed 
holonomies at the edges are $V(\vec x)$ and $V(\vec x + i)$.

In a recent paper~\cite{CDMP2} we 
have shown that QCD2 on a cylinder has a third order phase transition of 
the same type of that found by M.Douglas and V.Kazakov in the spherical 
geometry. Such a transition can be seen in the language of 2d QCD as 
an istanton driven delocalization of the trivial semicircle distribution 
of eigenvalues.
We will show in the next section that  in the present context of Finite 
Temperature LGT  this phase transition is responsible for the instability
of the weak coupling solution. 
Before proceeding to do that however, we shall end the present section 
by reviewing and completing the results of ref. ~\cite{CDMP2} on the 
Douglas-Kazakov phase transition on a cylinder.

\subsection{Douglas-Kazakov phase transition on a cylinder}

As  previously discussed  the Heat Kernel effective
 theory for the Polyakov loop  is
 given in terms of 
the partition function 
${\cal K}_2 ( g_1,g_2;\ca) \equiv {\cal K}_2(\phi,\theta;\ca)$ 
of QCD2 on a cylinder, with fixed holonomies (Polyakov loops) 
$g_1$ and $g_2$ (whose invariant angles are $\{\phi_i\}$ and 
$\{\theta_i\}$ 
respectively) at the two boundaries.

This partition function is the solution of the Heat Kernel equation
 on the
$SU(N)$ group manifold :
\eq
\left( N\frac{\partial}{\partial \ca} - \frac{1}{2}  {\cal J}^{-1}(\phi)   
\sum_i \frac{\partial^2}{\partial \phi_{i}^{2}}  {\cal J}(\phi)
- \frac{1}{24} N(N^2-1)
\right) {\cal K}_2 (\phi,\theta; \ca)  =  0 
\label{heateq}
\en
uniquely determined by the condition :
\eq
\lim_{\ca \rightarrow 0} {\cal K}_2(g_1,g_2;\ca) =
\hat{\delta}(g_1,g_2^{-1})
\label{deltain}
\en
where, in the notations of ref. \cite{CDMP1},
${\cal J}(\phi)$ is the Vandermonde determinant for a unitary matrix:
\eq
{\cal J}(\phi) = \prod_{i<j} 2 \sin \frac{\phi_i - \phi_j}{2}
\lbl{vandermonde}
\en
and $\hat{\delta}$ is the invariant delta function on the group
 manifold. 
An explicit form of ${\cal K}_2(\phi,\theta;\ca)$ 
is usually given by the standard character expansion :
\eq
{\cal K}_2 ( g_1,g_2;\ca)= \sum_r \exp 
\left[- \frac{\ca}{2N} C_r^{(2)}  \right] 
 \chi_r(\phi) \chi_r(\theta) 
\lbl{kercyl}
\en
As pointed out in \cite{CDMP1,Panz}, upon redefinition of the Kernel by
${\cal K}_2(\phi,\theta;\ca) \rightarrow \hat{{\cal K}}_2 = {\cal J}(\phi)
{\cal J}(\theta){\cal K}_2$, eq. (\ref{heateq}) becomes the (euclidean) free 
Schrodinger equation for $N$ fermions on the circle,
where $\ca$ plays the role of the (euclidean) time.
This means that, because of the condition (\ref{deltain}),
 we can interpret
${\cal K}_2(\phi,\theta;\ca)$ as the euclidean transition amplitude 
of this system of fermions
from the configuration $\{\phi_i\}$ at the zero time to the
 configuration 
$\{-\theta_i\}$ at the time $\ca$.

By using the results of ref. \cite{AIz} and \cite{CDMP1} the 
above expression can be rewritten as a sum over the $N$ integers $l_i$ 
labelling the co-root lattice of $SU(N)$:
\eqa
{\cal K}_2 ( g_1,g_2;\ca) &=& \left( \frac{N}{4 \pi} \right)^{1/2}
 \exp \left(\frac{\ca}{24} (N^2 - 1) \right)\sum_P 
\frac{(\frac{\ca}{N})^{(1-N)/2}}{{\cal J}(\theta)
{\cal J}(\phi)} 
\nonumber \\ 
& & (-1)^{\sigma(P)  + \frac{N(N-1)}{2} }
\sum_{\{l_{i}\}}
\exp \left[ - \frac{N}{2\ca} \sum_{i = 1}^N
\left( \phi_i + \theta_{P(i)} + 2 \pi l_i \right)^2     \right] 
\lbl{cylinst}
\ena
where $P$ denotes a permutation (of sign $\sigma(P)$) of the indices. 
The meaning of the modular transformation relating this 
last expression to the
character expansion is clear:
the integers labelling the unitary representations in the character
 expansion 
(\ref{kercyl}) correspond to discretized momenta of the fermions on
 the circle,
while eq. (\ref{cylinst}) gives the corresponding coordinate
 representation 
and the integers in the co-root lattice are interpreted as the
 winding numbers
of the fermions.

 The large $N$ behaviour of the partition function 
(\ref{kercyl}) in terms of the eigenvalues distributions $\rho_0(x)$ and
$\rho_1(x)$ corresponding respectively to $\{\phi_i\}$ and $\{\theta_i\}
$ can be obtained starting from the differential equation
 (\ref{heateq}) and then applying the same
procedure used by Matytsin in \cite{Maty} to find the large $N$ limit 
of the Itzykson-Zuber integral.
The time (area) evolution of the eigenvalue distribution $\rho(x)$ is 
given by a Das-Jevicki hamiltonian ~\cite{dasjev} 
\eq
H\left[ \rho(x),\Pi(x)\right] = {1\over 2}\int dx \rho(x) \left\{ \left(
{\partial \Pi(x) \over \partial x} \right)^2 - {\pi^2 \over 3} \rho^2(x) 
\right)
\label{dasjev} \en
where $\Pi(x)$ is the canonical momentum conjugate to $\rho(x)$.
The only difference with respect to ref. ~\cite{Maty} is that the 
density $\rho(x)$ is here a periodic function of period $2 \pi$. 
In terms of the complex quantity
\eq 
f(x,t) = {\partial \Pi(x,t) \over \partial x} + i \pi \rho(x,t)
\label{mah} \en
the equations of motion become
the Hopf equation of motion for an ideal fluid :
\eq
\frac{\partial f}{\partial t} + 
f \frac{\partial}{\partial x} f = 0 ,
\label{hopf}
\en
with the boundary conditions
\eq
\rho(x,t=0) = \pi \rho_0(x) ~,~ \rho(x,t=\ca) = \pi \rho_1(x)
\label{boundary}
\en

A solution to eq. (\ref{hopf}) can be obtained from the ansatz
\eqa
\rho(x, t)  =&   \frac{2}{\pi r^2(t)} \sqrt{r^2(t) - x^2} ,&~~~~~~
 |x|< r^2(t) \nonumber \\
\rho(x, t)  =0&, & r^2(t)<|x|<\pi
\label{Wigner}  
\ena

The ansatz (\ref{Wigner}) provides a solution of the Hopf equation 
if the time dependence of the radius $r$ of the distribution has 
the form
\eq
r(t)  =  2 \sqrt{\frac{(t+\alpha)(\beta - t)}{\alpha + \beta}} .
\label{gensol}
\en
The arbitrary  constants $\alpha,\beta$ are determined by the boundary 
conditions (\ref{boundary}). Naturally the initial and final 
distributions in (\ref{boundary}) have consistently to be semicircular 
with radia $r(0) = r_0 , r(\ca) = r_1$. 
Given $r_0$ and $r_1$, the radius of the distribution is determined at 
any section of the cylinder by eq.(\ref{gensol}).
However  because of the periodicity condition on the eigenvalue 
distribution 
the density (\ref{Wigner},\ref{gensol}) is a solution of the
 saddle-point equations
(\ref{hopf},\ref{boundary}) only if $r(t) < \pi$ for any $t$ on the
 trajectory.
For any given value of $r_0$ and $r_1$, the maximum value of $r(t)$ 
increases as the area ${\cal A}$ increases , so the solution
 (\ref{hopf},\ref{gensol})
is valid only if the area ${\cal A}$ is smaller than a critical value
${\cal A}_c$, where the maximum radius equals $\pi$ and the eigenvalues
 fill up
the whole circle. 
The critical value of $\ca$ at which such phase transition occurs can be 
easily calculated and 
is given by
\eq
{\cal A}_c =
\sqrt{ \frac{1}{2} \left( \pi^4 - \pi^2 \frac{r_0^2 + r_1^2}{2}
+ \sqrt{(\pi^4 - \pi^2 r_0 r_1)^2 - \pi^6 (r_0 - r_1)^2} \right)}
\label{critarea}
\en
Note that the partition function of QCD on a sphere is
a particular case corresponding to $r_0=r_1=0$. Indeed
in this case the critical value (\ref{critarea}) becomes 
 ${\cal A}_c = \pi^2$,
which is just the value found by Douglas and Kazakov \cite{dk}.
Hence the large $N$  phase transition  between the small area
(gaussian) phase and the large area phase is the generalization
to the cyilinder  of the 
Douglas-Kazakov phase transition on the sphere .

The meaning of the phase transition was discussed in \cite{CDMP2}:
in the gaussian phase the eigenvalues are confined in the $(-\pi,\pi)$ 
interval, namely the configurations where a fraction of the eigenvalues 
wind around their configuration space do not contribute in the large $N$ 
limit. Correspondingly all integers $l_i$ in eq. (\ref{cylinst}) can be 
set to zero.
Such topologically non trivial configurations contribute instead in the 
large $N$ limit of the large area phase.

The role of instantons in inducing the Douglas-Kazakov phase transition
on the sphere was further investigated by Gross and Matytsin \cite{GrMa}.

\section{Weak Coupling}

As a result of the approximation schemes discussed in the previous 
sections, we can now write an effective action 
for the Polyakov loop starting either from the Wilson action or 
from the Heat Kernel action. Moreover in the large $N$ limit, we can 
assume that the saddle point solution is translational invariant and, as 
a consequence, the action reduces to a one plaquette integral, both in 
the Wilson and in the Heat Kernel case. 
The partition functions in the two cases are given by
\eqa
Z_W & =& \int dV \prod_{i=1}^d dU_i e^{NJ
\sum_{i=1}^d~Re~Tr~U_iVU^\dagger_i V^\dagger} \nonumber \\
& =& \int dV \left[ \int dU e^{NJ~Re~Tr~UVU^\dagger V^\dagger}
\right]^d~~~,
\label{zetawilson}
\ena
and
\eq
Z_H = \int dV \left[ {\cal K}_2(V,V^{\dagger};1/{J^H}) \right]^d~~~,
\label{zetahk}
\en
where $d$ is the number of space dimension and 
${\cal K}_2(V,V^{\dagger};1/{J^H})$ is the Heat Kernel 
on a cylinder discussed in the 
previous section.
The  couplings $J$ and $J^H$ are related 
to the  coupling $\beta_t$ of the theory defined on a lattice 
with $N_t$ links in the time direction by eqs. (\ref{blarge},
\ref{bsmall}, \ref{bheatkernel}) and keeping into account the redefinition 
(\ref{jbeta}).
We shall study the phase diagrams resulting from
(\ref{zetawilson}) and (\ref{zetahk}) keeping in mind that, in the
region where the two actions describe the same physical system, 
the same critical phenomena should occur at values of $J$ and
$J^H$ corresponding to the same value of $\beta_t$.
On the other hand, we know that the rescaling of $J^H$
given by eq.(\ref{bheatkernel}) is exact after decoupling the
spatial plaquettes. Thus the comparison between the two partition 
functions can be used to improve eqs (\ref{blarge}) and (\ref{bsmall}).
In the case of eq.(\ref{blarge}), for instance, we will find that 
the two $J$'s must differ by a constant term
for the two theories to match in the weak coupling region.

\subsection{Weak coupling expansion for the Heat Kernel action}

The partition function (\ref{zetahk}) was studied in detail by Zarembo 
in \cite{zar}, and part of our analysis will overlap with his work.
We will, however, be able to interpret the critical phenomena occurring
in the weak coupling region in terms of the Douglas-Kazakov phase 
transition in the underlying QCD2 structure.
Let us first rewrite eq. (\ref{zetahk}) explicitly, as an integral over 
the invariant angles of $V$.
\eq
Z_H = \int \prod d\theta_i \left[ {\cal J}^2(\theta) \right]^{(1-d)}
\left[\sum_{k_i} \sum_P (-1)^{\sigma(P)} e^{-\frac{1}{2} N 
J^H \sum_i (\theta_i - \theta_{P(i)} + 2 \pi k_i)^2} 
\right]^d~~~.
\label{zetahk2}
\en
As discussed in the previous section, if $J^H$ is larger than a
critical value (to be determined), the contributions of the winding
configurations can be neglected in the large $N$ limit, and all $k_i$'s 
in eq.(\ref{zetahk2}) can be set to zero.
Moreover, for small $\theta$, we have
\eq
{\cal J}^2(\theta) = \Delta^2(\theta) e^{-
\frac{1}{12} N \sum_i \theta_i^2 + O(\theta^4)}~~~.
\label{Jdelta}
\en
Inserting these results in eq.(\ref{zetahk2}) we obtain
\eq
\int \prod_i d\theta_i \left[ \Delta^2(\theta) \right]^{(1 - d)} 
e^{-N [d J^H - \frac{d-1}{12}] \sum_i \theta_i^2 } 
\left[\sum_P (-1)^{\sigma(P)} e^{N J^H 
\theta_i \theta_{P(i)}} \right]^d~~~.
\label{zetahk3}
\en
Provided $J^H$ is larger than its critical value, the only
approximation needed to go from eq.(\ref{zetahk}) to eq.(\ref{zetahk3})
consists in neglecting $O(\theta^4)$ terms in (\ref{Jdelta}).
The r.h.s. of eq.(\ref{zetahk3}) is a Kazakov-Migdal model with quadratic
potential in the mean field approximation, which was solved exactly
in the large $N$ limit, for any number of space dimension, by Gross 
\cite{GrossKM}.
The eigenvalue distribution is a standard Wigner distribution
\eq
\rho(\theta) = \frac{2}{\pi r^2} \sqrt{r^2 - \theta^2}~~~,
\label{Wigner1}
\en
where the radius $r$ of the distribution is given by
\eq
r^2 = \frac{4(2 d-1)}{J^{H} (m^2 (d-1) + d \sqrt{ m^4 -
 4(2d-1)})}~~~.
\label{erre2}
\en
The parameter $m$ which appears in (\ref{erre2}) is the usual mass term 
of the Kazakov-Migdal model. Following the standard normalization we 
have that the coefficient of the quadratic term in $\theta$ in 
eq.(\ref{zetahk3}) must be $\frac{m^2}{2 J^H}$. This implies
\eq
m^2 = 2d-\frac{d-1}{6 J^H}~~~.
\label{emme2}
\en
In \cite{zar}, Zarembo argues that a phase
transition occurs at a value of $J^H$ equal to 
(or larger than) the one for which $r^2$ becomes
complex. Up to the effect of terms of higher order in $\theta$,
we can in fact determine the exact value of $J^H$
at which the phase transition occurs, and we will
further argue that the phase transition is of third order.
From the analysis of the previous section, we know  that the kernel
${\cal K}_2(V,V^{\dagger};1/{J^H})$ undergoes a phase transition of the
Douglas-Kazakov type when
\eq
\left(\frac{1}{J^H}\right)^2 = \pi^4 - \pi^2 r^2~~~,
\label{DKpoint}
\en
where $r^2$ is given by (\ref{erre2}).
The results for the critical value of $J^H$, and for the 
value $r_c$ of the radius 
at the critical point, are given for different space 
dimensions in Table (I). The critical values of $J^H$ obtained
by Zarembo are also listed for comparison. 

\vskip0.3cm
$$\vbox {\offinterlineskip
\halign  { \strut#& \vrule# \tabskip=.5cm plus1cm
& \hfil#\hfil 
& \vrule# & \hfil# \hfil
& \vrule# & \hfil# \hfil
& \vrule# & \hfil# \hfil &\vrule# \tabskip=0pt \cr \noalign {\hrule}
&& $d$ && $r_c$ && $J_{c}^H,~DK$ &&  $J_{c}^H,~Z$    &
\cr \noalign {\hrule}
&& $2$ && $2.96$  && $0.321$ && $0.311$ 
& \cr \noalign {\hrule}
&& $3$ && $2.80$ && $0.226$ && $0.218$ 
& \cr \noalign {\hrule}
&& $4$ && $2.66$ && $0.192$ && $0.184$
& \cr \noalign {\hrule}
&& $\infty$ && $0.$ && $1/\pi^2\sim0.101$ && 
$1/12\sim0.083$
& \cr \noalign {\hrule}
}}$$
\begin{center}
{\bf Tab.I.}{\it~~ Value of $J^H$ below which the Wigner distribution
becomes unstable, from the Douglas-Kazakov phase transition
(third column) and from ref \cite{zar}.
The radius of the distribution at the critical point is given 
in the second column.}
\end {center}
\vskip0.3cm

We observe that our values are consistently just slightly higher than 
Zarembo's, which clearly indicates that we are dealing with the same
instability and not with two different physical phenomena, and on
the other hand shows that the present one is a better approximation.
As a consistency check, we notice that the critical radius is less 
than $\pi$ for any number of dimensions, whereas if we take the 
solution of \cite{zar} we see that it exceeds $\pi$
for $d=2$ and $d=3$.
In spite of the small value of $J^H$ we expect the weak coupling solution
to be very reliable above the Douglas-Kazakov phase transition particularly 
for large $d$, where the small value of $r_c$ ensures that the quartic terms 
which  have been neglected in eq. (\ref{zetahk3}) are indeed small.

To sum up, we have established that the Wigner distribution of 
eigenvalues becomes unstable at a value of the coupling constant given 
by (\ref{DKpoint}), and the transition is driven by the winding modes 
in the configuration space of the eigenvalues.
This analogy with the Douglas-Kazakov phase transition on 
a sphere gives an almost 
compelling argument that the corresponding phase transition is of the 
third order.
The appearance at the
critical point of new classical trajectories corresponding to winding 
eigenvalues has presumably the effect of 
spreading the distribution, especially at the extremes.
On the other hand, for low dimensions, this transition occurs at a 
radius of the eigenvalue distribution very near to $\pi$. This probably 
means that the maximum corresponding to the broken, deconfined phase 
becomes unstable, and the distribution would collapse into the uniform 
one. However this does not mean that the Douglas-Kazakov transition
can be identified with the deconfinement transition. 
This is not the case for high dimensions, where the critical radius 
is small, approaching zero as $d$ increases. In this case the 
Douglas-Kazakov phase 
transition is a transition from a classical Wigner
distribution to another one, so far unknown, but still presumably
peaked around the origin.

\subsection{Weak coupling expansion for the Wilson action}

The partition function $Z_W$  of eq.(\ref{zetawilson}) can be written 
as an integral over the invariant angles of the Polyakov loop $V$,
\eq
Z_W = \int \prod_i d\theta_i {\cal J}^2(\theta)
\left( {\cal I}(\theta,J) \right)^d~~~,
\label{zetaW}
\en
where 
\eqa
{\cal I}(\theta,J)& = & \int [dU] e^{N J~Re~Tr~(U
 \cos\theta
U^{\dagger} \cos\theta + U \sin\theta U^{\dagger} \sin\theta) } \nonumber \\ & = &
 \int [dU] e^{N J \sum_{i,j} |U|_{ij}
 \cos(\theta_i - \theta_j)}~~~,
\label{calI}
\ena
and we denote by $\cos\theta$ and $\sin\theta$ diagonal matrices with
eigenvalues $\cos\theta_i$ and $\sin\theta_i$ respectively.
In the large $N$ limit this integral can be
evaluated as a sum of saddle point contributions \cite{koganetal}. 
The result is
\eq
{\cal I}(\theta,J) = \sum_P \frac{\exp \left[ N J  
\sum_i \cos(\theta_i - \theta_{P(i)}) \right]}{H_P(\theta)} 
(1 + O(\frac{1}{N J}))~~~,
\label{calIasimp}
\en
where
\eqa
H_P(\theta)& = &  \prod_{i<j}
 \Bigl[ (\cos\theta_i - \cos\theta_j)
 (\cos\theta_{P(i)} - \cos\theta_{P(j)})\nonumber \\& +& (\sin\theta_i
 - \sin\theta_j) (\sin\theta_{P(i)} - \sin\theta_{P(j)}) \Bigr]~~~.
\label{accapi}
\ena

The denominator $H_P(\theta)$ in eq. (\ref{calIasimp}) has a double 
pole whenever two $\theta$'s coincide, so it can be written 
as $\Delta^2(\theta) \exp \left( h_P(\theta) \right)$, and
$h_P(\theta)$ can be easily calculated for small $\theta$, with the
result 
\eq
h_P(\theta) = - \frac{N}{3} \sum_i \theta_i^2 + \frac{N}{4} 
\sum_i \theta_i \theta_{P(i)} + O(\theta^3)~~~.
\label{den}
\en
For small $\theta$, the partition function can thus be written as
\eq
Z_W = \int \prod_i d\theta_i \left[\Delta^2(\theta) \right]^{1-d} e^{-N
[d(J - \frac{1}{3}) + \frac{1}{12}] \sum_i \theta_i^2 }
 \left[ \sum_P (-1)^{\sigma(P)} e^{N (J - \frac{1}{4}) 
\sum_i \theta_i \theta_{P(i)}}\right]^d~~~.
\label{zetaW2}
\en
Once again, this is a Kazakov-Migdal model with quadratic
potential, and it coincides with (eq.(\ref{zetahk3})) if we identify
\eq
J = J^H + \frac{1}{4}~~~.
\label{betahkw}
\en

We have thus identified the leading correction to eq.
(\ref{blarge}) away from the continuum (weak coupling) limit: the two
lattice theories defined by the Wilson and the Heat Kernel 
actions begin to differ along the respective renormalization group 
trajectories by a constant shift of their couplings. As $J$ decreases
higher order terms in the angles $\theta$ will become
relevant and the two theories will become increasingly different.

As for the phase diagram, one cannot apply to the Wilson action 
the same precise estimate based on the Douglas-Kazakov phase transition,
but only the simpler argument given by Zarembo. 
This is due to the fact that the Douglas-Kazakov transition crucially 
depends on the precise form of the Heat Kernel action, and in particular 
on the winding numbers $k_i$ which are neglected in the Kazakov-Migdal 
approximation. In any case, as mentioned above,
the phase diagram of the two
theories is the same as long as the relation 
(\ref{betahkw}) is duly taken into account.

\section{Strong coupling}

While the inverse coupling $\beta_t/N^2$ in the lattice theory with $N_t$ 
links in the time direction becomes very large as we approach the 
continuum limit, the effective coupling $J$ in the reduced 
$N_t = 1$  theory (\ref{zetawilson}) remains relatively
small, typically of order $1/d$, even at temperatures 
approaching the deconfining transition. 
An expansion in powers of $J$ is therefore appropriate
to study the critical phenomena in the effective models for the 
Polyakov loop considered in the previous section.
 
As a preliminary step let us establish some notations. We are 
interested in the large $N$ limit, so the fundamental quantity is the  
distribution $\rho(\theta)$  of the eigenvalues of the Polyakov loop. 
It is convenient to expand $\rho(\theta)$ in its Fourier modes
\eq
\rho(\theta) = \frac{1}{2 \pi} \sum_{n=-\infty}^{\infty}
 \rho_n e^{i n \theta}
= \frac{1}{2 \pi} \sum_{n=-\infty}^{\infty} x_n 
e^{i \alpha_n + i n \theta}~~~,
\label{rhon}
\en 
where $\alpha_n \in (- \pi/2,\pi/2)$ is the
argument of $\rho_n$ modulo $\pi$, and $x_n$ coincides with
 the modulus of $\rho_n$ up to a sign
\footnote{It is convenient for the following discussion to restrict
the range for $\alpha_n$ and have $x_n$ taking also negative values}.
The reality of $\rho(\theta)$ requires $\rho_{-n} =
\rho_n^{*}$, while the normalization  of $\rho(\theta)$ to $1$ in the 
interval $(-\pi,\pi)$ fixes $\rho_0 = 1$.

The inverse formula
\eq
\rho_n = \int_{-\pi}^{\pi} \rho(\theta) e^{-i n \theta}
\label{rhoinv}
\en
shows that $\rho_n$ corresponds to the large $N$ limit of the 
loop winding $n$ times in the time-like direction
with the given eigenvalue distribution. 
In particular, $\rho_{\pm 1}$ corresponds
to the large $N$ limit of the Polyakov loop.

The $Z_N$ invariance of the effective theory
becomes, in the large $N$ limit, a $U(1)$ invariance under the shift 
$\theta \to \theta + \delta$, that is $\alpha_n \to \alpha_n + n 
\delta$. If this symmetry is unbroken the eigenvalue
distribution is simply given by $\rho(\theta) = {1 \over 2 \pi}$, 
namely $x_n = 0$ for $n \not = 0$. In the broken phase the $U(1)$ 
symmetry connects the different vacua. If we choose the vacuum 
peaked at $\theta = 0$, then the symmetry of the action for $\theta_i 
\to - \theta_i$ will force the vacuum
distribution $\rho(\theta)$ to be even in $\theta$, thus fixing all
$\alpha_n$ to zero. In this situation the eigenvalue distribution 
takes the form
\eq
\rho(\theta) = {1 \over 2 \pi} \left[ 1 + \sum_{n=1}^{\infty} 
x_n \cos (n \theta) \right]~~~,
\label{alpha0}
\en
with $x_n$ real.

The next ingredient is the integration measure
${\cal J}^2(\theta)$, which in the large $N$ limit can be expressed
as a function of the loop variables
$\rho_n$, as
\eq
{\cal J}^2(\theta) = \exp \left[\lim_{y \to 1} {N^2 \over 2} \int_{-\pi}^{\pi}
d\theta~ d\varphi~ \rho(\theta)~ \rho(\varphi)~ \log(1 - 
y\cos (\theta-\varphi)) \right]~~~,
\label{Jasimp}
\en
where the double integral, which would be
divergent at $\theta=\varphi$, has been regularized by the inclusion of
the parameter $y$, and terms in the exponent suppressed by powers of $N$
have been neglected.
Eq. (\ref{Jasimp}) can be  calculated by expanding in powers of $\theta$
and $\varphi$ and resumming the resulting expression. The result is
\eq
{\cal J}^2(\theta) = \exp \left[\lim_{y \to 1} N^2 \left( C_0(y) + 
\sum_{k=1}^{\infty} C_k(y) x_k^2 \right) \right]~~~,
\label{Jas2}
\en
where $C_0(y)$ is an irrelevant divergent expression and $C_k(y)$ is
given by
\eq
C_k(y) = {1 \over k} \left[ \frac{1 - \sqrt{1 - y^2}}{y} \right]^k~~~.
\label{ck}
\en
After removing the divergence, the limit $y \to 1$ can be taken,
and gives  
\eq
{\cal J}^2(\theta) = \exp \left[ N^2 \sum_{k=1}^{\infty} {1 \over k} x_k^2
                    \right]~~~.
\label{Jas3}
\en

Let us consider now the partition function $Z_W$, given in eq.
(\ref{zetaW}).
We are going to show that the integral ${\cal I}(\theta,J)$
can be written in the large $N$ limit as
\eq
{\cal I}(\theta,J) = e^{ N^2 \sum_{k=1}^{\infty} 
J^k F_k(x,\alpha)}~~~,
\label{calIas}
\en
where the functions $F_k(x,\alpha)$ can be shown to depend only on 
$x_i$ and $\alpha_i$ with $i \leq k$, and can be determined in 
principle by using Schwinger-Dyson equations.

Correspondingly, the large $N$ limit of the partition function $Z_W$ 
becomes
\eq
{1 \over N^2} \ln Z_W =   \sum_{k=1}^{\infty} \left( d~J^k 
F_k(x,\alpha) - {x_k^2 \over k} \right)~~~.
\label{zetaWas}
\en
It is already obvious from (\ref{zetaWas}) that, for small enough 
$J$, the free energy has a maximum for $x_k = 0$, that is for 
a uniform distribution of eigenvalues.
The first order approximation of (\ref{zetaWas}) is well known (see for 
example \cite{dp}), and is simply $F_1(x,\alpha) = x_1^2$.
The free energy to this order becomes
\eq
{1 \over N^2} \ln Z_W \approx  (J d - 1) x_1^2~~~,
\label{firstord}
\en
with all other $x_k$ set to zero. The phase structure of 
(\ref{firstord}) is very simple: for $J < {1 \over d}$ the 
maximum of the free energy occurs at $x_1=0$ and we are in the 
unbroken phase. Above the critical point $J ={1 \over d}$ we 
have instead $x_1 = 1/2$, which is the maximum
value allowed by the positivity condition on $\rho(\theta)$.
This is exactly the phase diagram of the large $N$ limit of the 
effective action
(\ref{seff}) discussed at the beginning of the introduction. In fact it 
is easy to see that at this first order approximation the two actions 
are actually equivalent. This allows to better understand in which sense 
we expect, going to higher order in eq.(\ref{zetaWas}),
 to improve the effective action (\ref{seff}).

Let us compare the first order expansion of the Wilson action with the
corresponding one for the Heat Kernel action $Z_{HK}$, given in eq. 
(\ref{zetahk}) and  (\ref{zetahk2}). 
The strong coupling expansion for the kernel on the
cylinder coincides with the character expansion, which is a power 
series in $e^{-{1 \over 2 J^H}}$. 
The truncation of the series to the first order corresponds to a
truncation of the character expansion to fundamental representation,
namely
\eqa
\sum_r \chi_r(\theta) \chi_r(-\theta) e^{-{C_r^{(2)} \over 2 N 
J^H}} & \approx & 1 + 2 \chi_f(\theta) \chi_f(-\theta) 
e^{- {1 \over 2 J^H}} \nonumber \\
& \approx & \exp \left[ N^2 2 e^{- {1 \over 2 J^H}} 
x_1^2 \right]~~~,
\label{hkfirstorder}
\ena
where in the last step we have used the fact that $\chi_f(\theta) = N 
x_1 e^{i \alpha_1}$. At first order in the strong coupling 
expansion $Z_W$ and $Z_{HK}$ coincide, provided we make the 
identification
\eq
J = 2 e^{- {1 \over 2 J^H}}~~~, 
\label{betarel}
\en
in agreement with eq.s (\ref{blarge}) and (\ref{bsmall}).
Indeed, we have just rederived, comparing the first order strong coupling
expansions of $Z_W$ and $Z_{HK}$, the well known rescaling properties 
of the coupling constant $J$.

\subsection{Strong coupling expansion for the Wilson action}

The functions $F_k(x,\alpha)$, which characterize the strong coupling 
expansion of $Z_W$, can be determined by solving Schwinger-Dyson type 
of equations. Let us consider the integral
\eq
{\cal I}(J;\Phi,\Psi) = \int [dU] e^{NJ Tr \left( U\Phi_1 U^{\dagger}
\Psi_1 + U \Phi_2 U^{\dagger} \Psi_2 \right)}~~~,
\label{int2}
\en
where $\Phi_{\alpha}$ and $\Psi_{\alpha}$ are hermitian matrices.
This integral is  more general than the one in eq.
(\ref{calI}), which is obtained from ${\cal I}(J;\Phi,\Psi)$
by imposing the constraints
\eq
\sum_{\alpha=1}^2 \Phi_{\alpha}^2 =\sum_{\alpha=1}^2 \Psi_{\alpha}^2 = 1
\label{constr}
\en
and by identifying $\Phi$ with $\Psi$.
Notice that, for $\Phi_2 = \Psi_2 = 0$, eq. (\ref{int2}) reduces to
the well known Itzykson-Zuber integral.
The Schwinger-Dyson equations for ${\cal I}(J;\Phi,\Psi)$ can 
easily be proved; they read
\eq
Tr \left[ {1 \over N} {\partial \over \partial \Phi_{\alpha_1}} {1 \over
N}{\partial \over \partial \Phi_{\alpha_2}} ..... {1 \over N} {\partial \over
\partial \Phi_{\alpha_s}} \right] {\cal I}(J;\Phi,\Psi) = J^s Tr \left(
\Psi_{\alpha_1} \Psi_{\alpha_2}...\Psi_{\alpha_s} \right) {\cal I}(J;
\Phi,\Psi)~~~.
\label{SchDys}
\en
These equation could, in principle, be solved order by order in $J$.
It is however more convenient to make use of the fact that ${\cal I}
(J;\Phi,\Psi)$ reduces  for $\Phi_2 = \Psi_2 = 0$ to the 
Itzykson-Zuber integral, whose ``strong coupling'' expansion has been 
explicitly derived in \cite{izhc} up to the eighth order in 
$J$. 
In fact it is not difficult to verify that, given a term in the 
expansion of the Itzykson-Zuber integral, containing only 
$\Phi_1$ and $\Psi_1$, there is a unique extension of it to include
the dependence on $\Phi_2$ and $\Psi_2$
which satisfies the following requirements:
i) it is invariant under the $O(2)$ symmetry  of the integral 
(\ref{int2}) and of the constraints (\ref{constr});
ii) it contains in each term $\Phi_i$ and $\Psi_i$ with the same power,
and thus is invariant when $\Phi_i$ and $\Psi_i$ are rescaled 
respectively by a factor $\lambda$ and ${1 \over \lambda}$, which is a
symmetry of the integral before imposing the constraints (\ref{constr}).
Using this property, it is easy to generalize the known results 
concerning the Itzykson-Zuber integral, and calculate the functions
$F_k(x,\alpha)$ appearing in eq. (\ref{calIas}). 
Up to $k=4$ they are given by

\eqa
F_1(x,\alpha)& = &x_1^2~~~, \label{f1} \\ &  & \nonumber \\
F_2(x,\alpha)& =  &{1 \over 4} \left( 1 - 2 x_1^2 + 2 x_1^4 + x_2^2 -
 2 x_1^2 x_2 \cos(2 \alpha_1 - \alpha_2) \right)~~~, \label{f2} \\
&  & \nonumber \\   
F_3(x,\alpha)& = &{ 1 \over 12} \Bigl( 3 x_1^2 - 12 x_1^4 + 16 x_1^6 + 
12 x_1^2 x_2^2 + x_3^2 + 6 x_1^2 x_2 \cos(2 \alpha_1 - \alpha_2) 
\nonumber \\ &  &
 - 24 x_1^4 x_2 \cos(2 \alpha_1 - \alpha_2) + 4 x_1^3 x_3
 \cos( 3 \alpha_1 - \alpha_3)  \nonumber \\ &  &- 6 x_1 x_2 x_3
\cos(\alpha_1 + \alpha_2 - \alpha_3 ) \Bigr)~~~,
\label{f3} \\
&  & \nonumber \\
F_4(x,\alpha)& = & {1 \over 32} \Bigl( -1 + 34 x_1^4 - 144 x_1^6 +
 192 x_1^8 - 40x_1^2 x_2^2 + 226 x_1^4 x_2^2  \nonumber \\
&  &+ 4 x_2^4 + 20 x_1^2 x_3^2 + x_4^2 + 34 x_1^4 x_2^2 \cos(4 \alpha_1 - 2
 \alpha_2) \nonumber \\
&  & - 12 x_1^2 x_2 \cos(2 \alpha_1 - \alpha_2)+
 152 x_1^4 x_2 \cos(2 \alpha_1 - \alpha_2) \nonumber \\ & &  
 - 384 x_1^6 x_2 \cos(2 \alpha_1 - \alpha_2)
 - 48 x_1^2 x_2^3 \cos(2 \alpha_1 - \alpha_2)
 \nonumber \\ &  & - 16 x_1^3 x_3 \cos( 3 \alpha_1 - \alpha_3) 
+ 80 x_1^5 x_3 \cos( 3 \alpha_1 - \alpha_3) \nonumber \\ & & 
 +8 x_1 x_2 x_3 \cos( \alpha_1 + \alpha_2 - \alpha_3 ) 
 - 120 x_1^3 x_2 x_3 \cos(\alpha_1 + \alpha_2 - \alpha_3 )
 \nonumber \\ & & +
 16 x_1 x_2^2 x_3 \cos( \alpha_1 - 2 \alpha_2 + \alpha_3)
- 10 x_1^4 x_4 \cos(4 \alpha_1 - \alpha_4) \nonumber \\ &  & 
 + 20 x_1^2 x_2 x_4 \cos(2 \alpha_1 + \alpha_2 - \alpha_4)
 - 4 x_2^2 x_4 \cos(2 \alpha_2 - \alpha_4)  \nonumber \\ & &
 - 8 x_1 x_3 x_4 \cos(\alpha_1 + \alpha_3 - \alpha_4) \Bigr)~~~.
\label{f4}
\ena

Notice the invariance of the functions $F_k(x,\alpha)$  under $ \alpha_k
\to \alpha_k + k \delta$, as a result of the $U(1)$ invariance
of the theory in the large $N$ limit.

We can now substitute the expressions (\ref{f1} - \ref{f4}) into 
(\ref{zetaWas}), and study the free energy up to fourth order in 
$J$. 
We have already noticed that at first order a phase transition
occurs at $J = {1 \over d}$. The higher order terms in
$J$ can be regarded then as higher order terms in  
${1 \over d}$, and the strong coupling expansion as a large $d$ 
expansion.
Consider for example the coefficient of the quadratic term in $x_1$.
The vanishing of this coefficient signals an
instability of the symmetric vacuum $x_k = 0$. This leads us to study 
the equation
\eq
d J - {d \over 2} J^2 + {d \over 4} J^3 
- 1 + O(J^5) = 0,
\label{vacinst}
\en
which can be solved order by order in ${1 \over d}$, leading to
\eq
J = {1 \over d} + {1 \over 2 d^2} + { 1 \over 4 d^3} + 
O({1 \over d^5})~~~.
\label{betainst}
\en
It is instructive to compare this result with the result obtained by 
Zarembo \cite{zar} using the Heat Kernel action. He finds that
the exact value at which the instability occurs is given by 
$J^H={1 \over 2 \log (2 d - 1)}$. 
When inserted in (\ref{betarel}) this expression
gives  $J = {1 \over d - 1/2}$, whose expansion in powers of 
$1/d$ coincides with the one found in (\ref{betainst}) up to the third 
order. This is more that what one would have naively expected, as eq. 
(\ref{betarel}) was derived by equating the expansions of 
the Wilson and the Heat Kernel actions only at first order.

While eq. (\ref{betainst}) gives an upper limit for the critical value 
of $J$, finding the precise value at which the deconfinement 
transition occurs, requires a more detailed analysis.                  
As discussed at the beginning of this section, when searching for the 
broken vacuum we can set all $\alpha_n$ to zero, and regard the free 
energy as a function of the $x_n$ only.
By using eq. (\ref{f1}  - \ref{f4}) we obtain
\eqa
{1 \over N^2} \ln Z_W & =& d J x_1^2 +                 
{d J^2 \over 4} \left( 1 - 2 x_1^2 + 2 x_1^4 + x_2^2 -
 2 x_1^2 x_2  \right) \nonumber \\
&+  &    
{d J^3 \over 12} \Bigl( 3 x_1^2 - 12 x_1^4 + 16 x_1^6 + 
12 x_1^2 x_2^2 + x_3^2    
\nonumber \\ &+&  6 x_1^2 x_2- 24 x_1^4 x_2  + 4 x_1^3 x_3 -
 6 x_1 x_2 x_3  \Bigr)
\nonumber \\
& + & { d J^4 \over 32} \Bigl( -1 + 34 x_1^4 - 144 x_1^6 +
 192 x_1^8 - 40x_1^2 x_2^2 \nonumber \\ &+& 226 x_1^4 x_2^2  + 4 x_2^4 +
 20 x_1^2 x_3^2 + x_4^2 + 34 x_1^4 x_2^2  \nonumber \\
&-  &  12 x_1^2 x_2 +
 152 x_1^4 x_2   
 - 384 x_1^6 x_2 
 - 48 x_1^2 x_2^3 
  - 16 x_1^3 x_3 \nonumber \\ 
&+& 80 x_1^5 x_3  
 +8 x_1 x_2 x_3  
 - 120 x_1^3 x_2 x_3 +
 16 x_1 x_2^2 x_3
  \nonumber \\ &  -& 10 x_1^4 x_4+
  20 x_1^2 x_2 x_4 - 4 x_2^2 x_4 
 - 8 x_1 x_3 x_4  \Bigr) \nonumber \\
&-&x_1^2 - {1 \over 2} x_2^2 -{ 1\over 3} x_3^2 - {1\over 4} x_4^2~~~.
\label{quart}
\ena
We must now look for the maximum of (\ref{quart}), 
within the domain where $\rho(\theta)$ is positive or zero for any 
$\theta$. The equations for such a domain can be obtained in principle 
by requiring that, for some $\theta_0$,
\eq  
\rho(\theta_0) = {d \over d \theta}\rho(\theta) |_{\theta=
\theta_0} = 0
\label{physreg}
\en
 and 
by eliminating $\theta_0$ from the equations. The resulting equations 
for the $x_n$'s give the boundaries of the physical region. 
In order to have an intuitive picture of how the deconfining transition 
takes place let us set $x_3=x_4=0$ in (\ref{quart}) and plot the free 
energy as a function of $x_1$ and $x_2$ at various values of the 
coupling $J$. This is shown in Fig.1, where a contour plot 
of the free energy is shown for $d=2$ and $J =
0.55,0.60,0.66$.

\iffigs
\begin{figure}
\null\vskip 0.3cm
\epsfxsize = 6.7truecm
\epsffile{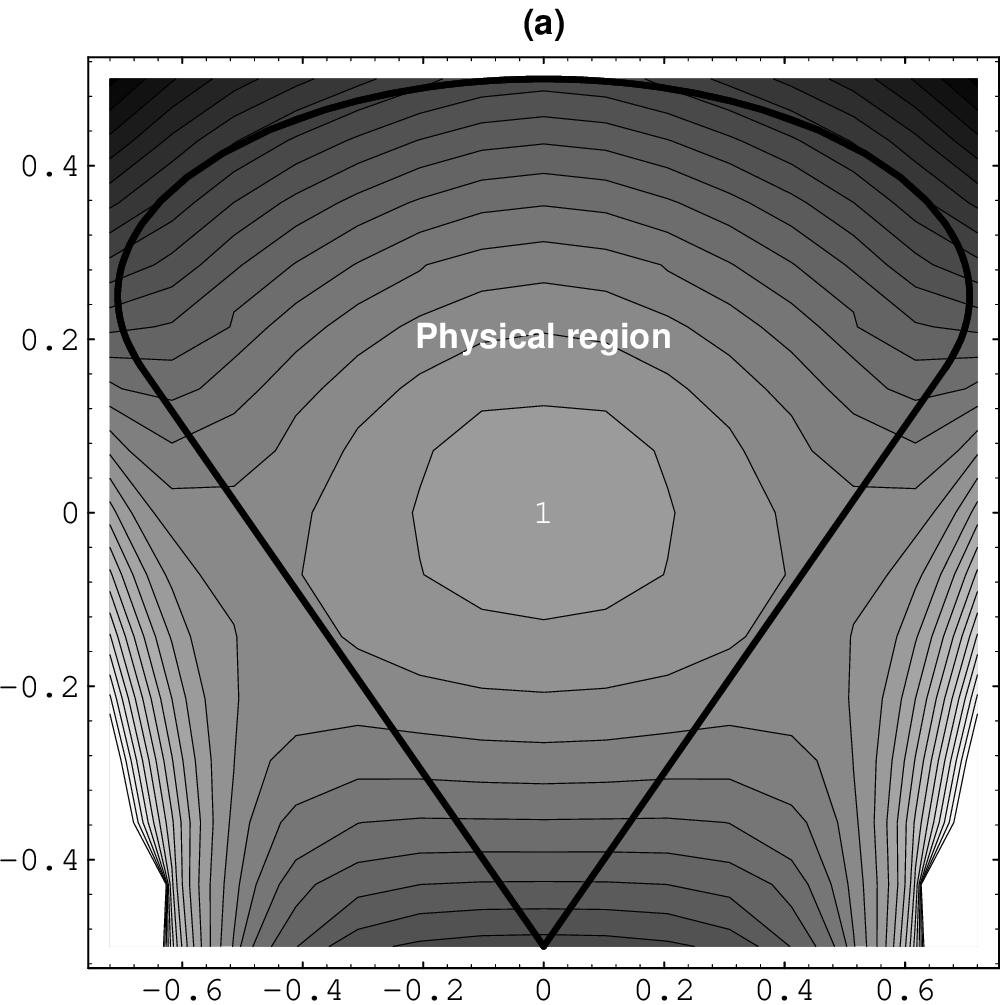}
\vskip -6.7truecm
\hskip 6.7truecm 
\epsfxsize = 6.7truecm
\epsffile{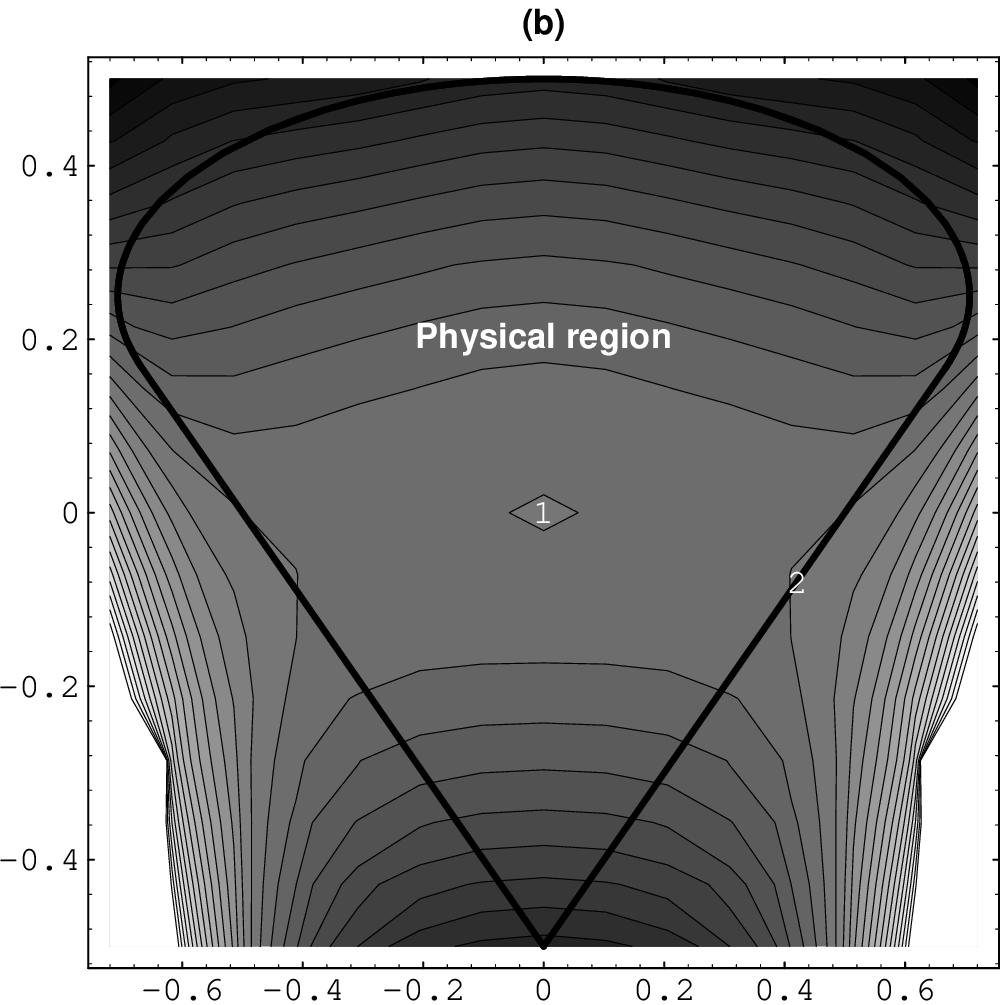}
\newline
\null
\hskip 3.2truecm 
\epsfxsize = 6.7truecm
\epsffile{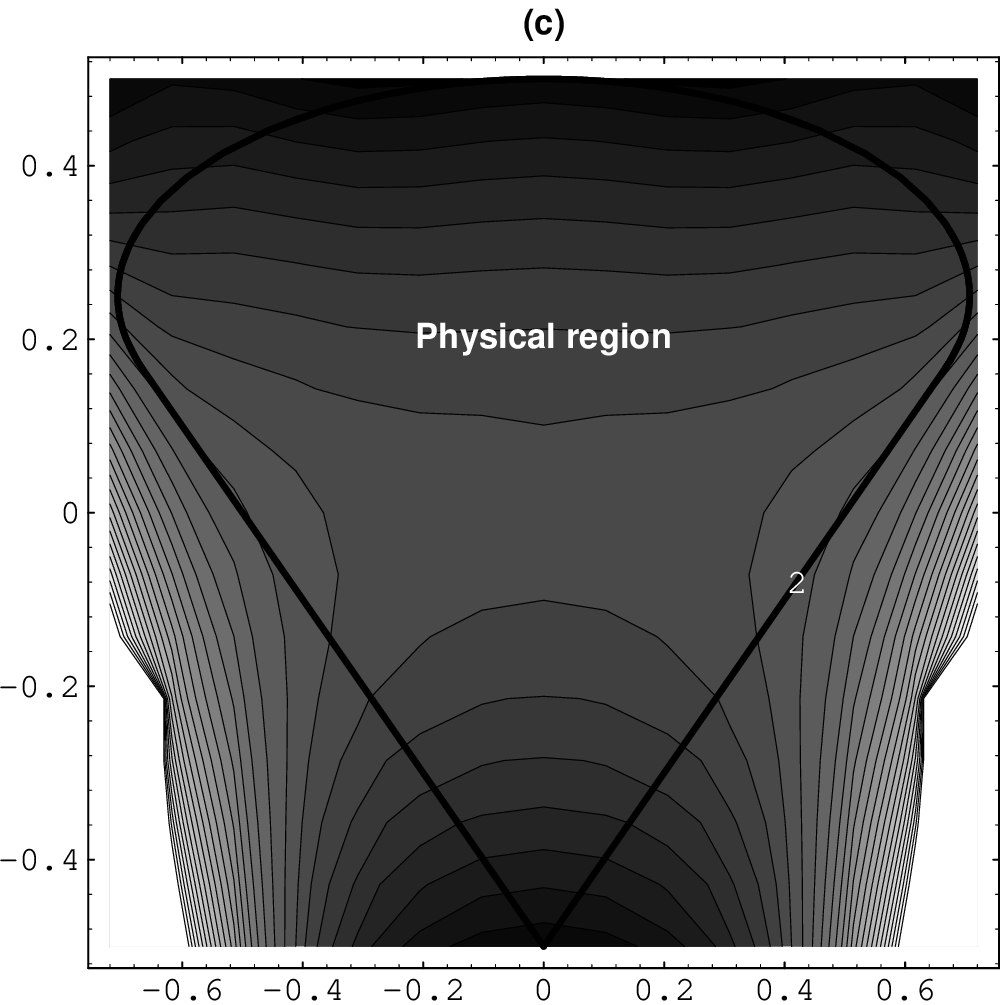}
\newline
{\bf Fig.1} {\it Contour plots of the free energy for: {\rm\bf
(a)}, $J = 0.55$; {\rm\bf (b)},$J = 0.61$; {\rm\bf (c)},
$J = 0.66$. The numbers {\rm\bf 1} and {\rm\bf 2} mark respectively
the presence of the symmetric maximum ($x_1 = x_2 = 0$) and the ``broken''
one. The thick line encloses the ``physical'' region} 
\label{contourplots}
\end{figure}
\fi

The physical region in the $(x_1,x_2)$ plane can be easily determined, 
and it is represented by the region inside the thick line. The straight 
edge on the left is given by the equation $x_1 - x_2 = 1/2$ and 
corresponds to density distributions vanishing at $\theta=\pi$, whereas 
the the straight edge on the opposite side corresponds to distributions 
vanishing at $\theta=0$.
The plot at $J= 0.55$ clearly shows the maximum at 
$x_1=x_2=0$: the system is in the unbroken phase dominated by a constant 
distribution of eigenvalues. In the next plot, at $J = 0.6$, 
a local maximum has appeared at the edge of the physical region, and it 
is becoming competitive with the unbroken maximum. Notice that there 
is a symmetry $x_1 \to -x_1$, so that a symmetric maximum appears on 
the other edge of the physical region. This symmetry is accidental, and
it is removed when $x_3$ and $x_4$ are switched on.
In the last plot, at $J = 0.66$, the maximum at $x_1=x_2=0$ 
has disappeared and the system is clearly in the broken phase.

When the dependence of (\ref{quart}) on $x_3$ and $x_4$ is taken into 
account, the boundary of the physical region is given by fourth order 
algebraic equations, together with the hyperplanes 
$x_1-x_2+x_3-x_4=1/2$ and $x_1+x_2+x_3+x_4=-1/2$. The phase transition 
occurs when the maximum value of the free energy on the boundary 
becomes larger than the value at the symmetric point. 
The critical point can be determined numerically, and it is given  
for different dimensions by the values listed in Table II.

\vskip0.3cm
$$\vbox {\offinterlineskip
\halign  { \strut#& \vrule# \tabskip=.5cm plus1cm
& \hfil#\hfil 
& \vrule# & \hfil# \hfil
& \vrule# & \hfil# \hfil
& \vrule# & \hfil# \hfil
& \vrule# & \hfil# \hfil
& \vrule# & \hfil# \hfil &\vrule# \tabskip=0pt \cr \noalign {\hrule}
&& $d$ && $J_{c}$ && $x_1$  && $x_2$ && $x_3$ && $x_4$  &
\cr \noalign {\hrule}
&& $2$ && $0.601$  && $0.50$ && $0.06$ && $-0.03$ && $-0.03$ 
& \cr \noalign {\hrule}
&& $3$ && $0.379$ && $0.52$ && $0.03$ && $-0.03$ && $-0.02$ 
& \cr \noalign {\hrule}
&& $4$ && $0.275$ && $0.53$ && $0.02$ && $-0.03$ && $-0.02$
& \cr \noalign {\hrule}
&& $100$ && $0.010$ && $0.49$ && $0.004$ && $0.004$ && $0.005$
& \cr \noalign {\hrule}
}}$$
\begin{center}
{\bf Tab.II.}{\it~~ Values of the critical coupling $J_c$ at the fourth 
order in the strong coupling expansion of the Wilson action. 
In the last four columns the 
corresponding values of $x_i,~~(i=1-4)$ are reported.}
\end {center}
\vskip0.3cm

The values of $J_c$ depends of course on the order at which the strong 
coupling expansion is truncated, the lowest order being, as already 
mentioned , $1/d$. The second and third order give for $d=2$ $J=0.66$ 
and $J=0.62$ respectively, showing that the value reported in Tab. II is 
likely to be an upper bound. Higher dimensions are much less sensitive, 
in agreement with the  idea that strong coupling expansions is also a 
large $d$ expansion.

It would be interesting then to discuss these results also
 from the point of view 
of the $1/d$ expansion, along the lines followed to study 
the instability of the symmetric vacuum. However, as the vacuum 
corresponding to the broken phase occurs on the boundary of the physical 
region, the constraints (\ref{physreg}) must be taken into account, 
giving an extra effective potential of order $1$ in $1/d$.
This means that already in the zeroth order of the expansion the Fourier 
components $x_n$ with $n > 1$ are not zero in general, and $x_1$ does 
not coincide with its na\"ive asymptotic value $x_1=0.5$ , as clearly 
indicated by Table II. The zero order approximation is already far 
from trivial when dealing with the broken maximum near the boundary, 
and we will not attempt its solution here.

\subsection{Strong coupling expansion for the Heat Kernel action}

We consider now the problem of obtaining a strong-coupling expansion 
for the effective theory of the Polyakov loops using the Heat 
Kernel action instead of the Wilson action. 

The starting point is naturally the character expansion
\eq
{\cal K}_2(g_1, g_2^{-1} ;1/J^H ) = \sum_R 
e^{- {C_R \over 2N J^H}} \chi_R (g_1) \chi_R (g_2^{-1})~~~.
\label{hkcyl1}
\en
Considering the translational invariance of the saddle point solution
for the effective 
theory, we are actually concerned with the case  
$g_1 = g_2 = V$, ${\rm Tr} V$ being the Polyakov loop.
However, we shall consider the more general case $g_1 \not = g_2$ 
and put $g_1 = g_2 = V$ only at the very end. 

The large $N$ limit of the kernel (\ref{hkcyl1}) was studied in Section 4
for large $\bhk$. The ``gaussian'' solution which arises in that 
situation is valid only down to a critical value of $\bhk$, 
where a third order Douglas-Kazakov phase transition takes place. 
It would be interesting to find out the exact expression in the
large $N$ limit of the kernel (\ref{hkcyl1}), also in the strong coupling
phase, extending the results of ref. \cite{dk} for the sphere
 to the cylinder.
 
Unfortunately, such an exact expression is not yet available. We shall 
instead give an expansion of the free energy $F(g_1,g_2^{-1};\bhk)$  
corresponding to the kernel (\ref{hkcyl1}) in powers of the relevant 
coupling, which in this case must be taken to be
\eq
\bw = 2 e^{-1/2J^H}~~~
\label{hkcyl1d}
\en
(notice that at the first order $\bw$ coincides with $J$).
Define the free energy in the large $N$ limit by
\eq
K_2(g_1,g_2^{-1};{1\over\bhk}) = e^{N^2 F(g_1,g_2^{-1};\bhk)}~~~,
\label{hkcyl1b}
\en
where as usual terms suppressed by powers of $N$ in the exponent will be 
neglected.
We can write then
\eq
F(g_1,g_2^{-1};\bhk) = 2~e^{-1/2\bhk} F_1^H(g_1,g_2^{-1};\bhk)~ 
 +~4~ e^{-1/\bhk} 
F_2^H(g_1,g_2^{-1};\bhk) + \ldots~~~,
\label {hkcyl1c}
\en
where the residual dependence on $\bhk$ in the individual terms is, as 
we shall see, polynomial in $1/\bhk$.

The expansion (\ref{hkcyl1c}) is of the same type of the one obtained 
when using the Wilson action; it is therefore what we need in order to 
compare the two cases. At lowest order, in fact, the two expansions 
coincide after the identification (\ref{hkcyl1d}).
 
We leave to  the Appendix  some technical remarks about the way to obtain 
such a strong coupling expansion, and about its meaning and its
potential connection with a string representation for the gauge theory. 
Presently, we quote the result of the expansion of the free energy 
$F(V,V^\dagger;\bhk)$ up to fourth order in $\bw$.
The result is expressed in terms of the Fourier modes of the density of 
eigenvalues $\rho (\theta)$, that is of the quantities $x_n$ and 
$\alpha_n$, introduced in the previous subsection.
We find, in the notations of the previous section,
\def\a#1{\alpha_{#1}} 
\eqa
F_1^H(x,\alpha;{J^H}) & = & x_1^2~~~, \label{hklong0} \\
F_2^H(x,\alpha;{J^H}) & = & 
{1\over 4} \biggl[-2 x_1^2 + {1\over 2{J^H}^2} x_1^4 +
x_2^2 - {2\over {J^H}} x_1^2 x_2 \cos (2\a 1 - \a 2)\biggr]~~~, \\
F_3^H(x,\alpha;{J^H}) & = &
{1\over 24}\biggl[6 x_1^2 - {6\over {J^H}^2} x_1^4 +{1\over 
{J^H}^4} x_1^6 + {12\over {J^H}^2} x_1^2 x_2^2 + 2 x_3^2  \nonumber \\
& & + {12\over {J^H}} x_1^3 x3 \cos (3\a 1 - \a 3) - {8\over {J^H}^3} 
x_1^4 x_2 \cos (2\a 1 - \a 2)  \nonumber \\
& & + {6\over {J^H}^2} x_1^3 x\cos (3\a 1 - \a 3) - {12 \over {J^H}} 
x_1 x_2 x_3 \cos (\a 1 + \a 2 - \a 3)\biggr]~~~, \\
F_4^H(x,\alpha;{J^H}) & = &
{1\over 16}\biggl[-2 x_1^2 + {4\over {J^H}^2} x_1^4 - {5\over 
2{J^H}^4} x_1^6 + {1\over 3{J^H}^6} x_1^8 - x_2^2  \nonumber \\
& & - {10\over {J^H}^2} x_1^2 x_2^2 + {10\over {J^H}^4} x_1^4 x_2^4 + 
{1\over {J^H}^2} x_2^4 + {6\over {J^H}^2} x_1^2 x_3^2 + {1\over 2} 
x_4^2  \nonumber \\
& & - {4\over {J^H}}x_1^2 x_2 \cos (2\a 1 - \a 2) + {38\over 3{J^H}^3}  
x_1^4 x_2 \cos (2\a 1 -\a 2)  \nonumber \\
& & -{4\over {J^H}^5} x_1^6 x_2 \cos (2\a 1 -\a 2)-{8\over {J^H}^3} 
x_1^2 x_2^3 \cos (2\a 1 -\a 2)  \nonumber \\
& & - {6\over {J^H}^2} x_1^3 x_3 \cos (3\a 1 -\a 3) + {9\over 2{J^H}^4} 
x_1^5 x_3 \cos (3\a 1 -\a 3)  \nonumber \\
& &{4\over {J^H}} x_1 x_2 x_3 \cos (\a 1 +\a 2 - \a 3) - 
{18\over {J^H}^3} x_1^3 x_2 x_3 \cos (\a 1 +\a 2 - \a 3)  \nonumber \\
& & + {6\over {J^H}^2} x_1 x_2^2 x_3 \cos (\a 1 - 2 \a 2 + \a 3)
- {8\over 3 {J^H}^3} x_1^4 x_4 \cos (4 \a 1 - \a 4)  \nonumber \\
& & + {8\over {J^H}^2} x_1^2 x_2 x_4 \cos (2 \a 1 +\a 2 - \a 4) -
{2\over {J^H}} x_2^2 x_4 \cos (2 \a 2 - \a 4)  \nonumber \\
& & {4\over {J^H}} x_1 x_3 x_4 \cos (\a 1 +\a 3 - \a 4) \biggr]~~~.
\label{hklong}
\ena

Some remarks are in order. First, as we mentioned above, each 
coefficient in the expansion of the free energy in powers of $\bw$ 
still contains powers of ${1 \over \bhk} $, that is of $\log \bw$. 
This is to be contrasted with the behaviour of the free energy 
associated with the Wilson action, where each term in the expansion is 
a function of $x_n$ and $\alpha_n$ only.
Second, the expansion (\ref{hkcyl1c}) agrees to all available orders 
with the result of \cite{zar}, that the symmetric vacuum becomes 
unstable at $\bhk = {1 \over 2 \log (2 d - 1)}$, or 
$\bw={1 \over d - 1/2}$.
This can be checked by looking at the coefficients  of $x_1^2$ in 
(\ref{hklong}), and following the arguments leading to (\ref{betainst}) 
in the previous subsection.

While the lowest order in $\bw$ of the expansion (\ref{hklong}) 
coincides with the corresponding one for the Wilson action, the higher 
orders are different, although the general structure remains the same.
It is interesting to see how the differences of the two models reflect 
in the critical values of $\bw$ at various dimensions.
The same type of numerical analysis that lead to Table
II using the Wilson action now gives the results listed in Table III.
The same critical values, in terms of the original Heat Kernel coupling 
$J^H$ are listed in the fourth column of Tab. IV.

\vskip0.3cm
$$\vbox {\offinterlineskip
\halign  { \strut#& \vrule# \tabskip=.5cm plus1cm
& \hfil#\hfil 
& \vrule# & \hfil# \hfil
& \vrule# & \hfil# \hfil
& \vrule# & \hfil# \hfil
& \vrule# & \hfil# \hfil
& \vrule# & \hfil# \hfil &\vrule# \tabskip=0pt \cr \noalign {\hrule}
&& $d$ && $\bw$ && $x_1$  && $x_2$ && $x_3$ && $x_4$  &
\cr \noalign {\hrule}
&& $2$ && $0.601$  && $0.41$ && $-0.13$ && $0.03$ && $0.07$ 
& \cr \noalign {\hrule}
&& $3$ && $0.339$ && $0.45$ && $-0.10$ && $-0.01$ && $0.05$ 
& \cr \noalign {\hrule}
&& $4$ && $0.238$ && $0.50$ && $-0.07$ && $-.05$ && $0.02$
& \cr \noalign {\hrule}
&& $100$ && $0.0096$ && $0.64$ && $0.07$ && $-0.10$ && $-0.03$
& \cr \noalign {\hrule}
}}$$
\begin{center}
{\bf Tab.III.}{\it~~ Same as Tab.II, but for the Heat Kernel action.
Notice the coupling $J_w$ is related to the Heat Kernel coupling through 
eq.(\ref{hkcyl1d}).}
\end {center}
\vskip0.3cm

The coincidence of the critical value of $\bw$ for $d=2$ with the one 
given in Table II is probably accidental. However, it can be seen from
eq. (\ref{hklong}) that some relevant coefficients in the first three 
terms of the expansion coincide with the corresponding ones for the 
Wilson action when $\bhk = 0.5$, that is when $\bw$ is around 0.75. 
This leads to expect a better agreement between the two theories in that 
region of $\bw$. In dimensions larger than $2$ the discrepancies are 
of the order of $10-15 \%$, and they probably give an idea of 
the dependence of the results on the chosen regularization.

\subsection{Summary of the results}

The results of both the weak and the strong coupling expansions are 
summarized in Tab.IV.

\vskip0.3cm
$$\vbox {\offinterlineskip
\halign  { \strut#& \vrule# \tabskip=.5cm plus1cm
& \hfil#\hfil 
& \vrule# & \hfil# \hfil
& \vrule# & \hfil# \hfil
& \vrule# & \hfil# \hfil
& \vrule# & \hfil# \hfil
& \vrule# & \hfil# \hfil
& \vrule# & \hfil# \hfil &\vrule# \tabskip=0pt \cr \noalign {\hrule}
&& $d$ && $J^H_{DK}$ && $J_{DK}$ && $J_c^H$ && $J_c$ 
&& $J^H_{s.v.i.}$ && $J_{s.v.i.}$   &
\cr \noalign {\hrule}
&& $2$ && $0.321$ && $0.561$ && $0.416$ && $0.601$ && $0.455$ && $0.667$
& \cr \noalign {\hrule}
&& $3$ && $0.226$ && $0.468$ && $0.282$ && $0.379$ && $0.310$ && $0.400$
& \cr \noalign {\hrule}
&& $4$ && $0.192$ && $0.434$ && $0.235$ && $0.275$ && $0.257$ && $0.286$
& \cr \noalign {\hrule}
&& $\infty$ && $\frac{1}{\pi^2}\sim0.101$ && $0.333$ && 
$\frac{1}{2\log(2d)}$ 
&& $\frac{1}{d}$ && $\frac{1}{2\log(2d)}$ && $\frac{1}{d}$ 
& \cr \noalign {\hrule}
}}$$
\begin{center}
{\bf Tab.IV.}{\it~~ Summary of our results. D.K refers to the Douglas-
Kazakov point, $s.v.i.$ to the symmetric vacuum instability point and 
$J_c,J^H_c$ are our best estimates for the deconfinement phase 
transition. See text for detailed 
explanation.}
\end {center}
\vskip0.3cm

In the second and third columns the weak coupling results are listed, 
namely the value of $J^H$ and $J$ for which the semicircular 
distribution becomes unstable (as discussed in sect.4), in the Heat 
Kernel and Wilson model respectively. 

In the two middle columns the best estimate for the phase transition, 
using the fourth order strong coupling expansion, is given for both 
models. In the last two columns the values of $J_H$ and $J$ at which the 
symmetric vacuum becomes unstable are given. They are obtained from the 
equations $J^H=1/(2\log(2d-1))$ and $J=1/(d-1/2)$, the former is an 
exact result~\cite{zar}, the latter is valid up to fourth order 
corrections in $1/d$ (see discussion in sect. 5.1).

These results are better understood with reference to Fig.2, where the 
relation between $J$ and $J^H$ is plotted in the weak coupling regime 
(curve (a)) and in the strong coupling regime (curve (b)). Notice the 
large overlapping region around $J^H=0.7$, where the strong and weak 
coupling definitions essentially coincide. 

The Douglas-Kazakov phase transition, marked by the points whose 
coordinates are in the second and third columns of Tab.IV, gives for each 
dimension $d$ the extreme of the curve (a) below which the weak coupling 
solution is not anymore valid. For very large $d$ this region goes as 
far down as $J^H=0.1$ into what one would naively expect to be a strong 
coupling regime. Indeed, as already remarked, it is really at large $d$ 
that the weak coupling solution can be already trusted just above the 
Douglas-Kazakov transition, as the radius of the Wigner distribution is 
then small, and the approximations made in neglecting quartic terms in 
$\theta's$ are reliable.

The instability of the symmetric vacuum on the other hand, marks
the end of the region where the strong coupling expansion can be relied 
upon. Beyond that point in fact the system is forced into a 
configuration where presumably all the Fourier modes $x_n$ of the 
eigenvalue distribution are excited (see discussion at the end of sect. 
5.1) and hence non perturbative effects in $J$ take place.
For lower dimensions (e.g. $d=2,3,4$) there is an interval of values of 
$J^H$ for which both curves are admitted. In that range the system has 
two classically stable configurations, one corresponding to the unbroken 
phase and one to the broken phase. 

As $J^H$ increases the system evolves along the curve (b), then {\it 
before} it reaches the point of instability for the symmetric vacuum it 
jumps onto the curve (a) {\it above} the Douglas-Kazakov point. The 
exact point of this transition has been estimated (earlier in this 
section) in the fourth order strong coupling expansion and it has been 
marked in Fig.2 for $d=2$ and $d=3$ with respectively a square and a 
triangle.

For large values of $d$ there is no overlap between the curves (a) and 
(b). This points to the existence of an intermediate phase where the 
symmetry is already broken, but the eigenvalue distribution is different 
from the semicircular one.

\iffigs
\begin{figure}
  \centerline{\epsfxsize=12cm\epsfbox{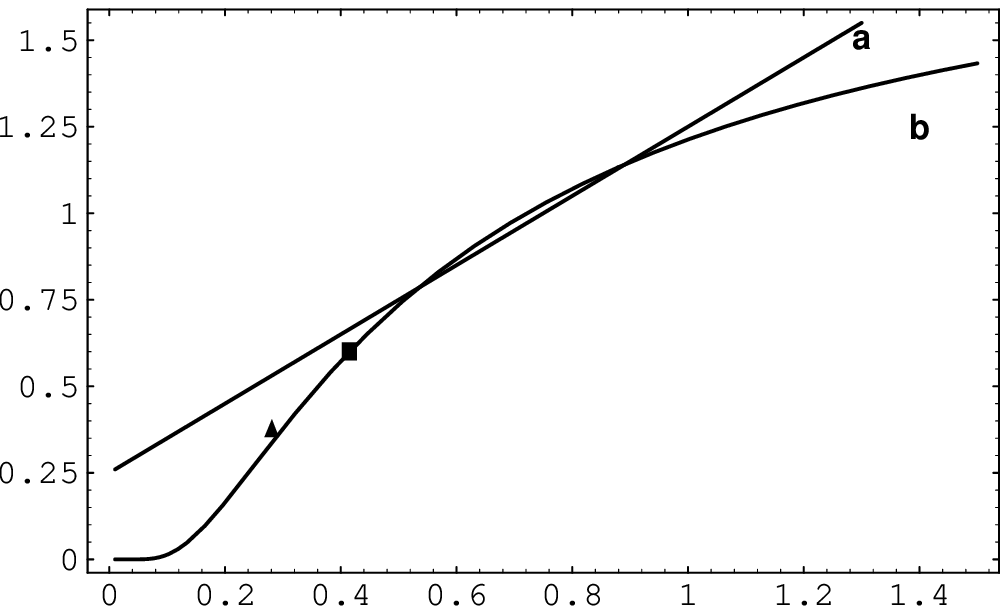}}
\vskip -3cm
\unitlength=1mm \begin{picture}(0,0)
\put(33,32){\makebox{$J^{H} + {1\over 4}$}}
\put(48,12){\makebox{$2 e^{- 1/ 2 J^{H}}$}}
\put(135,0){\makebox{$J^H$}}
\put(16,72){\makebox{$J$}}
\end{picture}
\vskip 1cm
{\bf Fig. 2.}{\it Redefinitions of the Heat Kernel inverse coupling
$J^{H}$ necessary to make contact with the Wilson case
{\bf (a)} in the weak coupling regime (large $J^{H}$),
and {\bf (b)} in the strong coupling one (small $J^{H}$).
In the region $0.4 <J^{H} < 1$ there is substantial
agreement of the two redefinitions.}
  \label{betaredef}
\end{figure}
\fi

\section{Conclusions}

Let us now try to compare our predictions on the critical deconfinement 
temperature with the known results obtained with Montecarlo simulations. 
Let us stress again that we must expect two kinds of systematic 
deviations. The first one is due to the large $N$ approximation. Indeed, 
the values that we have obtained in the large $N$ limit should be 
considered as an upper bound, and in the cases $N=2$ and $N=3$, 
in which most of the simulations have been performed, rather large 
deviations should be expected. However, this is not an unsolvable 
problem. The techniques described in this paper can in fact be used 
also at finite $N$, and work is in progress to do so.
Preliminary numerical tests 
show that the deviations in the predicted value of the critical 
temperature are exactly in the desired direction.

The second problem is that, in order to solve the model, 
we had to decouple space-like and time-like plaquettes in 
the original action. A direct consequence of this approximation is that 
the rule with which the coupling $\beta$ changes as a function of $N_t$ 
is essentially that of the strong coupling expansion. 
Once the spacelike plaquettes are decoupled, the theory becomes 
effectively two dimensional from the point of view of the strong 
coupling expansion, hence there is 
in principle no obstruction to extend the expansion to arbitrarly high 
values of $\beta$ and $N_t$, and the continuum limit can be taken. 
The scaling law of $N_t$ in this limit is completely fixed by
eq. (\ref{blarge}) if we use the Wilson regularization, or equivalently
by eq. (\ref{bheatkernel}) in the Heat Kernel case. We see that 
$\hat\beta/\beta=1/N_t$, or, equivalently, that
\eq
T_c=\frac{k}{\beta}~~~,
\label{scal}
\en
with k a constant to be determined.
As a direct consequence of the decoupling approximation, 
this scaling law holds in any dimensions and for any gauge group. 
The relevant point is that, in the (2+1) 
dimensional case, eq. (\ref{scal}) in fact
coincides with the correct scaling behaviour 
of the whole theory. This is probably due to the fact that in this 
case the decoupled spatial theory is a two dimensional gauge model, and,
as such, essentially trivial. 
Notice that a similarly 
good behaviour of the decoupling approximation in (2+1) dimensions was 
already noticed in~\cite{cd} in the complementary problem of describing 
the spacelike string tension. 

The second important point is that
 our value of the deconfinement temperature 
(which is effectively extracted as if we had $N_t=1$) 
must be, {\sl already at $N_t=1$}, in the weak coupling regime of the 
theory, unless we would have an additional uncertainty in making the 
extrapolation to $\beta \to \infty$. Remarkably enough, 
in $(2 + 1)$ dimensions the values that we find fulfills also this 
second condition, both in the Wilson and in the Heat Kernel case. Notice 
however that these values are
 at the border of this region, a fact that 
will require some further caution, and will be better discussed in the 
following subsection.

This must be contrasted with the (3+1) 
dimensional case. First, in the full theory a 
completely different scaling law is expected. Second, the critical 
point, at least in the Wilson case, is below the threshold of validity of 
the weak coupling expansion (see Tab. IV).
 These are actually two 
aspects of the same problem, and 
 they are warning us that the result obtained 
within our approach must be definitely considered as strong coupling
result. These considerations show that a more refined analysis,
for instance by 
using the techniques of the (hot, twisted) Eguchi-Kawai model, would be 
important in this case.

It should be noticed, however, that an incorrect scaling of
dimensional quantities does not necessarily imply a total loss of 
predictive power. In fact, as recently noticed in~\cite{fhk}, while
even the most precise and recent simulations of SU(2) and SU(3) 
models in (3+1) dimensions still show strong deviations from 
asymptotic scaling, scaling of dimensionless ratios is fulfilled even at 
small couplings, and very small values of $N_t$ (down to $N_t=3$).
This opens the way to a possible 
improvement in our predictions, if other dimensional quantities besides 
the critical temperature, say the 
string tension, could be evaluated, for instance in the framework 
outlined in~\cite{cd}. 

Let us now discuss, in turn, the $(2 + 1 )$ and the $(3 + 1)$ 
dimensional cases.

\subsection{2+1 dimensions}
In (2+1) dimensions our best prediction for the large $N$ value of the 
critical coupling $J_c=\hat\beta_c/N^2$ is $J_c=0.601$. Remarkably 
enough, this result follows both from the Wilson and the 
Heat Kernel regularizations, once the relation between Wilson and 
Heat Kernel couplings is used. 
The result should be considered 
as the fourth order improved version of the standard mean-field result
$J_c=1/d=0.5$. Comparing with the second and third order results we see 
that this should be considered as an upper bound, and that 
further orders would probably decrease the value.
Keeping in mind  the various factors of $N$, we obtain

\eq
\frac{T_c}{g^2}\equiv \frac{\beta_c}{N^2~N_t}=0.601~~~.
\en 

Looking at Tab. IV and Fig. 2 we see that this value is at the border of the 
weak coupling region, hence its extrapolation toward the continuum limit
$(N_t\to\infty, \beta_c\to\infty)$ could be affected by systematic 
errors. A simple way to estimate these possible deviations is to notice 
that, had we chosen from the beginning a Heat Kernel 
regularization for the whole theory, then, as discussed in the 
introduction, the scaling law eq. (\ref{scal}) would have been exact. 
In that case we could not have used eq. (\ref{betarel})
 to map the Heat Kernel 
coupling to the Wilson one; instead, we would have had to keep the true 
Heat Kernel critical coupling, which turns out to be (see sect.5 and 
Tab.IV) 
$J_c^{H}=0.416$.
Since the simulations with which we shall compare our results were 
performed with the Wilson action we shall use $J_c=0.601$ in our 
comparison, but 
the difference between $J_c$ and $J_c^{H}$ gives us an idea of size of 
the possible systematic errors involved in the continuum limit.

To the best of our knowledge, Montecarlo estimates of the 
deconfinement temperature in (2+1) dimensions only exist for the $N=2$ and 
$N=3$ models. 

In the case of SU(2), very precise and careful estimates of the 
critical temperature exist~\cite{t2} for the lattice sizes 2, 3, 4. The 
values are reported in Tab. V, together with an estimate of the 
continuum limit $(\beta\to\infty)$, using only the leading $O(1/\beta)$
corrections (see~\cite{t2} for details). While the deviations from 
scaling are still rather strong in the $T_c/g^2$ data they are, as 
expected, much more under control for the dimensionless ratio 
$T_c/\sqrt{\sigma}$, whose continuum limit requires the inclusion of
only $O(1/\beta^2)$ corrections.

\vskip0.3cm
$$\vbox {\offinterlineskip
\halign  { \strut#& \vrule# \tabskip=.5cm plus1cm
& \hfil#\hfil 
& \vrule# & \hfil# \hfil
& \vrule# & \hfil# \hfil
& \vrule# & \hfil# \hfil &\vrule# \tabskip=0pt \cr \noalign {\hrule}
&& $N$ && $N_t$ && $T_c/g^2$  &&  $T_c/\sqrt(\sigma)$  &
\cr \noalign {\hrule}
&& $2$ && $2$ && $\sim 0.433$ && $\sim 1.015$ & \cr \noalign {\hrule}
&& $2$ && $3$ && $\sim 0.417$ && $\sim 1.063$ & \cr \noalign {\hrule}
&& $2$ && $4$ && $\sim 0.412$ && $\sim 1.088$  & \cr \noalign {\hrule}
&& $2$ && $\infty$ &&  $0.385(10)$ && $1.121(8)$ & \cr \noalign {\hrule}
&& $3$ && $2$ &&  $\sim 0.454$ && $$ & \cr \noalign {\hrule}
&& $\infty~\cite{cd}$ &&$$ &&  $0.477$ && $0.977$ & \cr \noalign {\hrule}
&& $\infty$, $J_c$ &&$$ &&  $0.601$ && $$ & \cr \noalign {\hrule}
&& $\infty$, $J_c^{H}$&& $$ &&  $0.416$ && $$ & \cr \noalign {\hrule}
}}$$
\begin{center}
{\bf Tab.V.}{\it~~ The deconfinement temperature $T_c$ and 
the dimensionless ratio $T_c/\sqrt{\sigma}$ as  a function of the 
lattice size in the t direction $N_t$ in the (2+1) dimensional SU(2) 
and SU(3) LGT. Taken from~\cite{t2} and~\cite{ctdw}. In the sixth row we 
report the estimate of ref.~\cite{cd}, while our predictions are 
reported in the last two rows.} 
\end {center}
\vskip0.3cm

In the case of SU(3) there is, to the best of our knowledge, only one 
numerical result. It is the critical coupling for the $N_t=2$ lattice, 
which turns out to be $\beta_c(N_t=2)\sim 8.17$~\cite{ctdw}. This 
corresponds to $T_c/g^2\sim 0.454$. No extrapolation to the continuum 
limit is possible in this case, however it is interesting to see that the 
value is definitely higher than those of the $N=2$ case. This is a 
signature of the fact that as $N$ increases the $T_c/g^2$ values moves 
toward our large $N$ limit. 

Finally, let us  mention  the result obtained in~\cite{cd} 
(also reported in Tab. V) by using as additional input the 
dimensionless ratio $\sigma=\pi T_c^2/3$, which comes from the 
effective string approach to the interquark potential in LGT's.
The resulting prediction for the  critical temperature is 
\eq
T_c=\frac{3(N^2-1)}{2\pi\beta}~~~,
\en
which means, in the large N limit, $T_c/g^2=0.477$, a result which 
again is in the same region of our large $N$ estimates.

\subsection{3+1 dimensions}

In this case we can test our results with a much larger sample of data.
The main problem is obviously that the scaling laws (namely the 
dependence of $\beta_c$ on the value of $N_t$) of our simplified model 
and of the whole theory  are completely 
different. This essentially means that, due to the decoupling of the 
space-like plaquettes (and to the fact that in this case the purely 
spacelike theory is  non trivial), the lattice spacing in the time 
direction, with which we measure $N_t$ in our decoupled model,
is related in a non trivial way to the lattice spacing of the starting 
theory. 
The only way to use our results is then to compare our decoupled 
theory and the Montecarlo simulations of the 
full theory for small values of $N_t$, where $J_c$ is small and we can 
trust our strong coupling expansion. In this case we must definitely 
study the Wilson action only, and no universality argument can be used.
The scaling law is given by eq. (\ref{bsmall})

Our best estimate for the critical coupling in $(3 + 1)$ dimensions 
is $J_c=0.378$. 
Very precise data on SU(2) and SU(3) in the range $N_t=2-16$ can be found 
in~\cite{fhk}. 
Their values in terms of the $\Lambda$ scale
defined as 
\eq
\Lambda=\left(\frac{24\pi^2\beta}{11N^2}\right)^{(51/121)}
exp(-\frac{12\pi^2\beta}{11N^2})
\label{scalqcd}
\en
together with the corresponding values of $J_c\equiv\beta_c/N^2$
are shown in Tables VIa and VIb. 

\vskip0.3cm
$$\vbox {\offinterlineskip
\halign  { \strut#& \vrule# \tabskip=.5cm plus1cm
& \hfil#\hfil 
& \vrule# & \hfil# \hfil
& \vrule# & \hfil# \hfil &\vrule# \tabskip=0pt \cr \noalign {\hrule}
&& $N_t$ && $\beta_c/N^2$ && $T_c/\Lambda$  &
\cr \noalign {\hrule}
&& $2$ && $0.4700(8)$ && $29.7(2)$ & \cr \noalign {\hrule}
&& $3$ && $0.5442(8)$ && $41.4(3)$ & \cr \noalign {\hrule}
&& $4$ && $0.5746(2)$ && $42.1(1)$ & \cr \noalign {\hrule}
&& $5$ && $0.5932(11)$ && $40.6(5)$ & \cr \noalign {\hrule}
&& $6$ && $0.6066(8)$ && $38.7(3)$ & \cr \noalign {\hrule}
&& $8$ && $0.6279(10)$ && $36.0(4)$ & \cr \noalign {\hrule}
&& $16$ && $0.6849(25)$ && $32.0(8)$ & \cr \noalign {\hrule}
}}$$
\begin{center}
{\bf Tab.VIa.}{\it~~ The critical coupling $\beta_c/N^2$ and the 
corresponding deconfinement temperature $T_c/\Lambda$ 
 as  a function of the 
lattice size in the t direction $N_t$ in the (3+1) dimensional SU(2) 
LGT. The data are taken from~\cite{fhk}.}
\end {center}
\vskip0.3cm

\vskip0.3cm
$$\vbox {\offinterlineskip
\halign  { \strut#& \vrule# \tabskip=.5cm plus1cm
& \hfil#\hfil 
& \vrule# & \hfil# \hfil
& \vrule# & \hfil# \hfil &\vrule# \tabskip=0pt \cr \noalign {\hrule}
&& $N_t$ && $\beta_c/N^2$ && $T_c/\Lambda$  &
\cr \noalign {\hrule}
&& $2$ && $\sim 0.568$ && $\sim 78.8$ & \cr \noalign {\hrule}
&& $3$ && $0.6167(11)$ && $85.7(1.0)$ & \cr \noalign {\hrule}
&& $4$ && $0.63250(2)$ && $75.41(2)$ & \cr \noalign {\hrule}
&& $6$ && $0.65490(6)$ && $63.05(4)$ & \cr \noalign {\hrule}
&& $8$ && $0.667(3)$ && $53.34(1.5)$ & \cr \noalign {\hrule}
&& $10$ && $0.6844(8)$ && $51.05(40)$ & \cr \noalign {\hrule}
&& $12$ && $0.6964(13)$ && $48.05(65)$ & \cr \noalign {\hrule}
&& $14$ && $0.7092(11)$ && $46.9(5)$ & \cr \noalign {\hrule}
&& $16$ && $0.716(6)$ && $44.3(2.5)$ & \cr \noalign {\hrule}
}}$$
\begin{center}
{\bf Tab.VIb.}{\it~~ The critical coupling $\beta_c/N^2$ and the 
corresponding deconfinement temperature $T_c/\Lambda$ 
 as  a function of the 
lattice size in the t direction $N_t$ in the (3+1) dimensional SU(3) 
LGT. The data for $N_t \ge 3$ are taken from~\cite{fhk}, the
single one at 
$N_t=2$ is taken from~\cite{ksw}.}
\end {center}
\vskip0.3cm

The values of $J_c$ from Tables VIa and VIb corresponding to small 
values of $N_t$, together with some results for a large value of $N$ 
$(N=81)$ obtained through Montecarlo simulations~\cite{daskogut}
 of the hot twisted Eguchi-Kawai model are compared 
with our predictions in Table VII. Our values are obtained using 
eq.(\ref{bsmall}) and the value $(N_t=1,~~J_c=0.378)$ as input.

\vskip0.3cm
$$\vbox {\offinterlineskip
\halign  { \strut#& \vrule# \tabskip=.5cm plus1cm
& \hfil#\hfil 
& \vrule# & \hfil# \hfil
& \vrule# & \hfil# \hfil
& \vrule# & \hfil# \hfil
& \vrule# & \hfil# \hfil &\vrule# \tabskip=0pt \cr \noalign {\hrule}
&& $N_t$ && $N=2$ && $N=3$  && $N=81$ && $N=\infty$  &
\cr \noalign {\hrule}
&& $2$ && $0.4700(8)$  && $\sim 0.568$ && $$ && $0.615$ 
& \cr \noalign {\hrule}
&& $3$ && $0.5442(8)$ && $0.6167(11)$  && $$ && $0.723$ 
& \cr \noalign {\hrule}
&& $4$ && $0.5746(2)$ && $0.63250(2)$  && $$ && $0.784$
& \cr \noalign {\hrule}
&& $5$ && $0.5932(11)$ && $$ && $$ &&  $0.823$ 
& \cr \noalign {\hrule}
&& $5.14$ && $$ && $$ &&  $0.695(5)$ && $0.828$ 
& \cr \noalign {\hrule}
&& $6$ && $0.6066(8)$ && $0.65490(6)$  && $0.705(5)$ && $0.850$ 
& \cr \noalign {\hrule}
}}$$
\begin{center}
{\bf Tab.VII.}{\it~~ The critical coupling $\beta_c/N^2$ as a function 
of $N_t$ and $N$. In the last column our predictions, according to the 
scaling law eq.(\ref{bsmall}) and $\beta_c/N^2=0.378$ for $N_t=1$.
The data for $N=2$ and $3$ are taken form Tab. VIa and VIb. Those for 
$N=81$ are taken from~\cite{daskogut}.}
\end {center}
\vskip0.3cm

We may observe that
\begin{description}
\item{a]} 
the scaling law eq.(\ref{bsmall}) is definitely different from the 
scaling behaviour eq.(\ref{scalqcd}). However
 the Montecarlo estimates of the critical couplings
seem to be  still far from the asymptotic
scaling region and the values of $T_c/\Lambda$  keep 
decreasing as $\beta_c$ and $N_t$ increase. 
It is interesting to notice that the critical couplings for small values 
of $N_t$
have a dependence from $N_t$ which is not too different from our strong 
coupling expectation eq.(\ref{bsmall}).
\item{b]} As expected, the best agreement between data and predictions
is obtained for our lowest value of $N_t$: $N_t = 2$.
\item{c]} 
At fixed $N_t$, as  $N$ increases also $J_c$ systematically increases, 
and our large $N$ estimates seem to be the upper limit of this behaviour.
\end{description}

\vskip 0.5cm

As we have seen, our results, obtained analytically in the large $N$
limit, show a more than qualitative agreement with Montecarlo 
simulations performed on the full theory; having used two different
lattice regularizations gives us some control on the systematic errors 
associated with the continuum limit; the uncertainties due to the large 
$N$ approximation and to the decoupling of the space-like degrees of 
freedom can be further reduced by studying the model at finite $N$, and 
by using the Eguchi-Kawai approach. We believe that a combination of 
these methods may allow in future further improvements in the analytic 
determination of the deconfinement critical temperature. 

\def\a#1{\alpha_{#1}}
\def\b#1{\beta_{#1}}
\def\r#1{r_{#1}}
\def\s#1{s_{#1}}
\def\k2{{\cal K}_2}

\vskip 1cm
\appendix{\Large {\bf{Appendix}}}
\label{hkscapp}
\vskip 0.5cm
\renewcommand{\theequation}{A.\arabic{equation}}
\setcounter{equation}{0}

We are interested in the large-N limit of the Heat Kernel on the cylinder 
\begin{equation}
{\cal K}_2(g_1, g_2^{-1} \hskip 2pt ;\hskip 2pt \bhk ) = \sum_R 
e^{- {C^{(2)}_R \over 2 \bhk N}} \chi_R (g_1) \chi_R (g_2^{-1})
\label{aphkcyl1}
\end{equation}
In sec. (3) we utilized this kernel as a building block
for the effective 
theory of Polyakov loops. We were therefore with the case  
$g_1 = g_2 = V$, ${\rm Tr} V$ being the Polyakov loop and  
we gave in eq.(\ref{hklong0}-\ref{hklong}) an expansion of the free 
energy corresponding 
to the partition function (\ref{aphkcyl1}) in powers of $2e^{-1{1\over 2\bhk}}$, 
up to $4^{\rm th}$ order.

We consider here the general case, sketching  the procedure to obtain 
explicitly such an expansion and discussing a little its meaning.
 
The kernel (\ref{aphkcyl1}) is a particular case of the QCD2 partition 
function on manifolds ${\cal M}_{G,b}$ of genus $G$ with $b$ boundary 
components, the case  ${\cal M}_{0,2}$. 
As it is well known after \cite{GT} the 
character expansion of the QCD2 partition function on ${\cal M}_{G,b}$ 
can be interpreted as a 
weighted sum over a suitable set of maps from world-sheets of all 
possible genera to ${\cal M}_{G,b}$, that is, as a string theory.

The ``string coupling constant'' of this 
peculiar string theory is $1/N$ - there's a factor of $N^{2 - 
2g}$ in front of the contributions of the maps from genus $g$ 
world-sheet. Since we confine ourself to the $N=\infty$ limit, i.e. we 
do not care about $O({1\over N})$ corrections in eq. (\ref{hkcyl1b}),
the expansion we are looking for corresponds to 
consider just the string tree level ($ g = 0$) contributions.

The string interpretation of the QCD2 partition function arises technically 
limiting  the sum over the 
$SU(N)$ representations to the representations obtained by symmetrizing  
tensor products of the fundamental representation and of its conjugate.
As shown by the Douglas-Kazakov exact solution at $N =\infty$ in the case of 
the sphere this is the correct way to carry out the sum in the strong-coupling 
(small-area) regime.
We too will therefore sum in eq. (\ref{aphkcyl1}) over these 
representations. 

They can be in general labeled by means of two sets of integers 
$\{l\}$ and $\{m\}$
satisfying
\begin{equation}
l_1\geq l_2\geq \ldots \geq l_p\geq 0 \geq
m_1\geq m_2\geq \ldots \geq m_q \hskip 5pt
\label{hkcyl1e}
\end{equation}
The $\{l\}$ and the negative of the $\{m\}$ identify two Young tableaux, 
$L$ and $M$, of order $l = \sum_i l_i$ and $m = - \sum_i m_i$; 
the representation is then called a ``composite representation'' in 
\cite{GT} and denoted as $(\bar M L)$. Its order can be defined as $l + m$.
When the $\{m\}$ are all zero, the Casimir is given by
\begin{equation}
C^{(2)}_L = l N + l + \sum_i l_i (l_i - 2 i) - 
{l^2 \over N}\hskip 5pt .
\label{hkcyl1f}
\end{equation} 
For the representation $(\bar M L)$ the Casimir is 
\begin{equation}
C^{(2)}_{(\bar M L)} = C^{(2)}_L + C^{(2)}_M + 2 {l m\over N}.
\label{hkcyl1g}
\end{equation}
Equations (\ref{hkcyl1f},\ref{hkcyl1g}) are the key point allowing to 
obtain an expansion in powers of $e^{-1/2\bhk}$ for the kernel 
(\ref{aphkcyl1}). Indeed  eq. (\ref{aphkcyl1}) becomes
\begin{eqnarray}
{\cal K}_2(g_1, g_2^{-1} \hskip 2pt ;\hskip 2pt \bhk ) &=& \sum_{l,m = 
0}^{\infty} e^{-{l + m\over 2\bhk}} \sum_{L\in Y_l} \sum_{M\in Y_m}
\chi_{(\bar M L)} (g_1)\chi_{(\bar M L)} (g_2^{-1}) \cdot\nonumber\\
&&
e^{-{1\over 2\bhk} 
\left[({l + \sum_i l_i(l_i - 2 i)\over N} - {l^2\over N^2}) + (l 
\leftrightarrow m) + 2 {l m\over N^2}\right] } .
\label{hkcyl1i}
\end{eqnarray}
$Y_l$ denotes here the set of Young tableaux made of $l$ boxes. 

The characters $\chi_{(\bar M L)}(g_i)$ can be handled in the large-N limit 
rewriting them in terms of the quantities 
\begin{equation}
\rho_k = {1\over N} {\rm Tr}\hskip 2pt g_1^k\hskip 0.2cm ;\hskip 0.2cm
\sigma_k = {1\over N} {\rm Tr}\hskip 2pt g_2^k
\label{hkcyl1l}
\end{equation}
and of their conjugates $\rho_k^* (= {1\over N} {\rm Tr}\hskip 2pt 
(g_1^{\dagger})^k)$ and $\sigma_k^*$. 
These quantities remain finite as $N \rightarrow \infty$, turning into the
Fourier modes of the eigenvalue distributions $\rho(\theta)$ 
and $\sigma(\phi)$ corresponding to $g_1$ or $g_2$:
\begin{equation}
\rho_k = \int\hskip -3pt d\theta\hskip 3pt \rho(\theta) e^{- i k\theta}
\label{hkcyl1m}
\end{equation}
This rewriting of the characters is accomplished by means of  the 
Frobenius representation, extended to accomodate the case of 
composite representations $(\bar M L)$ \cite{GT},\cite{Bars}.

Having made explicit all the dependences from $N$ it is possible to expand
explicitly eq. (\ref{hkcyl1i}) up to a certain power (call it $P$) of 
$e^{-{1\over2\bhk}}$, which means to consider all the representations 
$(\bar M L)$ such that $l + m \leq P$, to write the resulting
expression as the exponential of a free energy and to neglect the subleading 
terms as $N\rightarrow\infty$ in the latter, obtaining
finally an expression of the form (\ref{hkcyl1c}). The expansion of the 
free energy can of course be obtained  by subtracting from the expansion 
(\ref{hkcyl1i}) the ``disconnected'' contributions (for instance at 
order $e^{-{1\over\bhk}}$ the contributions of products two rep.s of order 
$l + m = 1$). 

Expanding the last exponentials in eq. (\ref{hkcyl1i}) on gets power series in 
$1/\bhk N$ and $1/\bhk N^2$. However, when extracting the $O(N^2)$ terms 
only in the free energy, at fixed order $l + m$  only a polynomial in 
$1/\bhk$ survive
\footnote{This fact has a nice interpretation in the 
QCD2 string language; it is related to the existence of a limiting value for  
the Euler characteristic of the covering space at fixed characteristic 
of the target.}. The resulting expression makes thus sense as a strong 
coupling expansion.

Now we give the free energy, eq. (\ref{hkcyl1c}) up to $3^{\rm rd}$ 
order. We use the trigonometric notation for the complex modes of the 
eigenvalue distributions $\rho$ and $\sigma$:
\eq
\rho_k = r_k e^{i\alpha_k}\hskip 0.2cm ;\hskip 0.2cm
\sigma_k = s_k e^{i\gamma_k}\hskip 0.2cm ;\hskip 0.2cm
(r_k, s_k\in{\bf R}) .
\en
\def\b{{J^H}}
\def\bw{e^{-1/2\b}}
\def\a#1{\alpha_{#1}}
\def\g#1{\gamma_{#1}}
\begin{eqnarray}
&&F(g_1,g_2^{-1};\b) = \nonumber\\
&&\bw r_1 s_1 \cos (\a 1 - \g 1) + {(\bw)^2\over 4} \biggl[- r_1^2 - s_1 ^2
- {1\over\b} r_2 s_1^2 \cos (\a 2 - 2 \g 1) \nonumber\\
&&- {1\over\b} r_1^2 s_2 \cos (2 \a 1 - \g 2) - {2\over \b} r_1^2 s_1^2 
\cos (\a 1 - \g 1)\ + r_2 s_2 \cos^2 (\a 2 - \g 2)\nonumber\\
&& +{(1 + 4\b)\over 2\b^2} r_1^2 s_1^2 \cos \bigl(2(\a 1 - \g 1)\bigr)\biggr]
\nonumber\\
&&+ {(\bw)^3\over 24}\biggl[{-4- 6\b\over\b^3} r_1 r_2 s_1^3 
\cos (\a 1 + \a 2 - 3\g 1) +{3\over\b^2} r_3 s_1^3 \cos (\a 3 - 3\g 1) 
\nonumber\\
&& + \bigl(6 r_1 s_1 - {3\over \b^2} (r_1^3 s_1 + r_1 s_1^3) -
{1 +\b\over\b^3} r_1^3 s_1^3 \bigr) \cos(\a 1 - \g 1)\nonumber\\
&& + {1+ 3\b + 3\b^2\over\b^4}r_1^3 s_1^3 \cos \bigl(3(\a 1 - \g 1)\bigr)
+ \bigl({6\over\b} r_1 r_2 s_1 +{6\over\b^2} r_1 r_2 s_1^3\bigr)
\cos (\a 1 - \a 2 + \g 1) \nonumber\\
&&+ {-4 -6\b\over\b^3}r_1^3 s_1 s_2 \cos(3\a 1 - \g 1 - \g 2) +
{12 + 6\b\over\b^2} r_1 r_2 s_1 s_2 \cos(\a 1 + \a 2 - \g 1 - \g 2)
\nonumber\\
&&-{6\over\b} r_3 s_1 s_2 \cos (\a 3 - \g 1 - \g 2)  +\bigl( 
{6\over\b}
r_1 s_1 s_2 +{6\over\b^2} r_1^3 s_1 s_2 \bigr) \cos (\a 1 +\g 1 - \g 2)
\nonumber\\
&&-{6\over\b} r_1 r_2 s_1 s_2 \cos(\a 1 - \a 2 - \g 1 + \g 2)
+{3\over\b^2} r_1^3 s_3 \cos (3\a 1 - \g 3)\nonumber\\
&& -{6\over\b} r_1 r_2 s_3 \cos (\a 1 + \a 2 - \g 3) 
+ 2 r_3 s_3 \cos (\a 3 - \g 3)
\label{hkcyl10}
\end{eqnarray}
In the case $g_1 = g_2$ the free energy up to $4^{\rm th}$ order has 
been given in eq. (\ref{hklong}).

\vskip 1cm
\centerline{{\bf  Acknowledgements}}
\vskip 0.5cm
We thank F.Gliozzi for many helpful discussions.
A.D. would like to thank Yu. Makeenko for an interesting discussion.


\begin{thebibliography}{99}


\bibitem{sy} B. Svetitsky and L.Yaffe,{\it Nucl. Phys.}{\bf B210} (1982)
423.

\bibitem{og} M. Ogilvie \PRL{52} (1984) 1369.

\bibitem{djk} J.M.Drouffe, J.Jurkiewicz and A.Krzywicki, \PR{D29} (1984) 
2982.

\bibitem{dp} P.H.Damgaard and A.Patkos, \PL{B172} (1986) 369.

\bibitem{gnr} A.Gocksch, F.Neri and P.Rossi, \PL{B143} (1984) 207.

\bibitem{ps} J.Polonyi and K.Szlachanyi, \PL{B110} (1982) 395.

\bibitem{gk} F.Green and F.Karsch, \NP{B238} (1984) 297.

\bibitem{dh} P.H.Damgaard and M.Hasenbusch, preprint CERN-TH-7222/94, 
April 1994. 

\bibitem{heatkernel} J.-M.Drouffe,\PR{D18} (1978) 1174;

P.Menotti and E.Onofri, \NP{B190} (1981) 288.

\bibitem{zar} K. Zarembo, preprint SMI-94-7 (May, 1994).

\bibitem{dk} M.R.Douglas and V.A.Kazakov, \PL{B319} (1993) 219.

\bibitem{kad} A. Migdal, Sov. Phys. JETP {\bf 42} (1976) 413, 743.
L.P. Kadanoff, Ann. Phys. (N.Y.) {\bf 100}, (1976) 359, and Rev. Mod. 
Phys. {\bf 49} (1977) 267.

\bibitem{KazMig} V.A.Kazakov and A.A.Migdal, \NP{B397} (1993)
214.

\bibitem{CDPht} M.Caselle, A.D'Adda and S.Panzeri, \PL{B302} (1993) 80.

\bibitem{cd} M.Caselle A.D'Adda preprint DFTT 8/94, To be published on 
Nucl. Phys. B.

\bibitem{t2} M.Teper, \PL{B311} (1993) 223, \PL{B313} (1993) 417. 

\bibitem{izhc} Harish-Chandra, Amer.J.Math. {\bf 79} (1957) 87;

C.Itzykson and J.B.Zuber, J.Math.Phys. {\bf 21} (1980) 411;

M.L.Mehta, Comm.Math.Phys. {\bf 79} (1981) 327.

\bibitem{bu} T.Banks and A.Ukawa, \NP{ B255} (1983) 145.

\bibitem{bhdg} J.Bartholomew, D.Hochberg, P.H.Damgaard and M.Gross \PL{B
133} (1983) 218.

\bibitem{CDMP2} M.Caselle,A.D'Adda,L.Magnea and S.Panzeri, preprint 
DFTT 50/93, to appear in ``Proceedings of the Summer School in High Energy 
Physics and Cosmology'', Trieste 1993.

\bibitem{CDMP1} M.Caselle, A.D'Adda, L.Magnea and S.Panzeri, \NP{B416}
 (1994) 751.

\bibitem{Panz} S.Panzeri, Mod. Phys. Lett. {\bf A8} (1993) 3201.

\bibitem{AIz} D.Altschuler and C.Itzykson, Ann. Inst. H. Poincar\'e,
Phys. Th\'eor. {\bf 54} (1991) 1.


\bibitem{Maty} A.Matytsin, \NP{B411} (1994) 805.

\bibitem{dasjev} S.R.Das and A.Jevicki, Mod. Phys. Lett.
 {\bf A5}(1990) 1639.


\bibitem{GrMa} D.Gross and A. Matytsin, preprint PUPT-1459 (April 1994).

\bibitem{GrossKM}  D.Gross, \PL{B293} (1992) 181.

\bibitem{koganetal} I.I.Kogan  {\it et al.}, \NP{B395} (1993) 547. 

\bibitem{fhk} J.Fingberg, U.Heller and F.Karsch, \NP{B392} (1993) 493.


\bibitem{ctdw} J. Christensen, G. Thorleifsson, P.H. Damgaard and 
J.F. Wheater, \NP{B374} (1992) 225.


\bibitem{ksw} J.B.Kogut {\sl et al.}, \PRL{50} (1983) 393.

\bibitem{daskogut} S.Das and J.B.Kogut, \PR{D31} (1985) 2704.

\bibitem{GT} D.Gross, \NP{B400} (1993) 161;

D.Gross and W.Taylor, \NP{B400} (1993) 181; {\bf B403} (1993) 395.

\bibitem{Bars} I.Bars, J. Math. Phys. {\bf 21} (1980) 2678.



\end{thebibliography}
\end{document}